**Supershear Rupture Indicator in Near-fault Particle Motion**

Suli Yao[1,2], Hongfeng Yang[1,3,4,5*], Harsha S. Bhat[2], and Hideo Aochi[2,6]

[1]Department of Earth and Environmental Sciences, The Chinese University of Hong Kong, Hong Kong, China

[2]Laboratoire de Géologie, École Normale Supérieure, CNRS-UMR 8538, PSL Research University, Paris, France

[3]Shenzhen Research Institute, The Chinese University of Hong Kong, Shenzhen, Guangdong, China

[4]Institute of Environment, Energy and Sustainability, The Chinese University of Hong Kong, Hong Kong, China

[5]Institute of Space and Earth Information Science, The Chinese University of Hong Kong, Hong Kong, China

[6]Bureau de Recherches Géologiques et Minières (BRGM), 45100 Orléans, France

* Corresponding author: hyang@cuhk.edu.hk

## Abstract

Earthquake rupture propagation speed is an essential source parameter that fundamentally controls hazard and risk. In this paper, we develop and demonstrate the capability of a single near-fault seismic record in delineating earthquake rupture speed by inspecting displacement particle motion. We apply this method to near-fault strong-motion data in M7+ strike-slip earthquakes and identify diverse particle-motion signatures associated with subshear rupture, supershear rupture, supershear transition, initial rupture expansion, and potential multiple rupture fronts. This study highlights the value of near-fault observations for determining rupture speed.

## Introduction

Understanding earthquake rupture propagation speed is essential as it largely governs seismic hazard intensity. Shallow crustal earthquakes with high rupture speeds are likely to generate intense ground shaking due to their rapid energy release. Earthquake rupture speed can sometimes exceed the shear wave speed (supershear) (*1*, *2*), as demonstrated in the 1979 Imperial Valley Mw 6.5 earthquake (*3*), the 1999 Mw 7.6 İzmit, Türkiye earthquake (*4*, *5*), and the 2001 Kunlunshan Mw 7.8 earthquake (*6*). In a supershear rupture, the coherent S waves produce high-amplitude Mach cones that propagate with weak spatial attenuation (*7*), posing threats to surrounding regions even at considerable distances. Therefore, rapid determination of the rupture speed after the occurrence of a major earthquake is important for loss estimation and emergency response planning. Moreover, understanding factors that control rupture speed and propagation characteristics holds important implications for earthquake source physics. Currently, rupture speeds are predominantly determined using network-based methods, such as back projection (*8*) and finite-fault inversion (*9*), which have been well developed and can, to some degree, be operated automatically with high efficiency. However, the results of these methods sometimes lead to controversy because obtaining reliable results requires careful data compilation and model parametrization. It is also well recognized that identifying non-persistent supershear rupture propagation is very challenging for these traditional methods. Thus, additional robust and independent constraints on the rupture speed are still needed.

Strong-motion seismic or high-rate GNSS stations within several kilometers of the fault (Fig. 1A) can be used to directly track rupture propagation including stopping phases (*10*, *11*), as they record high-amplitude (on the order of m/s) velocity pulses upon the arrival of the rupture front (*12*). Given a number of stations distributed along the fault, rupture propagation can be delineated through tracking continuous velocity pulse records, as demonstrated in the 2023 Mw 7.8 Kahramanmaraş earthquake (*13*). However, in most cases, only a single station record is available for a given event, such as in the 2002 Denali Mw 7.8 earthquake (*14*) and the more recent 2025 Mw 7.7 Myanmar earthquake (*15*, *16*). Nevertheless, even a single record can provide valuable constraints through pinpointing the local rupture time and yielding the average rupture speed for persisting rupture (*5*, *17*). Furthermore, as illustrated in theoretical models and several cases, the shape and amplitude of the velocity pulses can distinguish between subshear and supershear ruptures (*10*, *18–22*).

Even though several stations record relatively simple velocity pulses that have been well studied (*22*, *23*), most near-fault records exhibit waveform complexities that remain incompletely understood. Multiple long-period (<0.5 Hz) velocity pulses may be associated with varying off-fault distances, fault bends, or multiple slip episodes (*13*), whereas the origin of high-frequency wiggles is intrinsically difficult to identify (*24*).

These complexities obscure the rupture speed information encoded in the waveforms. Consequently, previous analyses of velocity pulses, such as phase picking or waveform fitting, have generally been conducted after applying bandpass or lowpass filters.

Instead of near-fault waveforms of the velocity pulses, here we focus on the displacement particle motion and seek a simple and robust approach to identify rupture speeds. In kinematic rupture simulations in half-space and 1D layered domains, the particle motion of displacement has been found to differ between end-member supershear and subshear cases (*25*). However, these models did not consider realistic complexities such as fault bends and transient supershear. Moreover, whether near-fault data in nature are consistent with such model predictions remains unknown. In recent years, the amount of near-fault data has grown (Fig. 1A), providing an opportunity to investigate and validate theoretical predictions. In this study, we demonstrate the particle motion patterns in heterogeneous ruptures using kinematic and dynamic rupture models, and then identify various signatures in the real data associated with subshear rupture, supershear rupture, initial rupture expansion, and supershear transition.

## Typical particle motions in subshear and supershear strike-slip ruptures

In this section, we revisit the signatures of particle motion associated with subshear and supershear strike-slip ruptures using kinematic and dynamic rupture models. In the kinematic models, we assume uniform right-lateral strike slip of 2 m on a vertical planar fault with a total length of 60 km and a depth extent of 10 km (Fig. 2A; see details in Materials and Methods). We place a near-fault station at the along-strike center, 1 km from the fault. The rupture propagates unilaterally with a vertical rupture front. As the rupture propagates, the receiving azimuth and the distance to the source evolve at the station, leading to the rotation of motion polarization. The contribution of the preceding rupture is dominantly in the fault-normal (FN) component, the local rupture (within 10 km of the station) contributes large fault-parallel (FP) motion, and the contribution of the receding rupture is limited (Fig. 2, A and B). For different rupture speeds, the radiation emitted from different positions arrives at the station with different relative timings, leading to different particle motion trends. For a subshear rupture, the pulses generated by the preceding rupture travel faster than the advancing rupture front (Fig. 2B), leading to an FN-FP-FN motion pattern (Fig. 2, C and D). In contrast, for a supershear rupture, the local rupture contribution arrives before that of the preceding rupture (Fig. 2B), producing an FP-FN motion pattern (Fig. 2, C and D).

We also simulate particle motions through dynamic rupture modelling (Figs. 3, S1, and S2; see details in Materials and Methods) with a 1-D layered velocity structure (Table S1) using PyLith, a finite-element tool (*26*, *27*). We start the earthquake as a subshear rupture and then stimulate a sustained supershear rupture by adding a high-shear-stress asperity (Fig. S1). Similar to the kinematic models, we observe an FN-FP-FN motion pattern in the sustained subshear stage and an FP-FN motion pattern in the sustained supershear stage (Fig. 3A).

The sensitivity of the particle motion trend to the rupture speed decreases with the off-fault distance. For stations within 3 km of the fault, the particle motion can differentiate the rupture speed along the fault segment within 20 km of the station (Figs. S3 and S4). In the following analysis, we therefore only consider stations within 3 km of the fault.

**Particle motions in heterogeneous ruptures**

Unlike the uniform rupture presented above, ruptures of natural earthquakes typically involve complexities associated with rupture initiation and termination, heterogeneous slip, rupture acceleration and deceleration, and complex fault geometry. In this section, we examine the response of particle motion to these heterogeneities and derive criteria for applying the method.

In the initial stage, the rupture nucleates at depth and expands circularly or elliptically. For nearby stations, the rupture acts as a local rupture, and regardless of the rupture speed, the dominant contribution is in the FP direction owing to the receiving azimuth, as illustrated by both the kinematic and dynamic models (Figs. 3A and S5). As the rupture propagates, the subshear or supershear features build up over ~5-10 km (Figs. 3 and S5). Close to the rupture termination (within ~5 km), the synthetics exhibit significantly oblique final displacements owing to the uneven slip distribution on the two sides of the station (Figs. 3E and S5). Beyond the rupture extent, the particle motion is dominated by FN motion associated with the S-wave stopping phase.

To account for heterogeneous slip, we test kinematic models with stochastic slip distributions (Fig. S6; see details in Materials and Methods), which yield nearly identical particle motions to those of the uniform-slip model, consistent with the results in the previous study (*25*). Adding a large asperity can induce oblique static displacements near its edges, but the overall particle-motion trend is preserved (Fig. S6). We also test the influence of the slip pulse duration (1 to 3 s) and find it to be minor (Fig. S7). Slip heterogeneities and pulse duration mostly influence the high-frequency energy, whereas the particle-motion trend is controlled by the low-frequency signal.

We further test the influence of transitions between subshear and supershear speeds (Fig. S8). Once a transition occurs, the particle-motion trend evolves over ~10 km along strike, and it responds faster to a supershear-to-subshear transition than the opposite. In the dynamic model, where the transition is smoother, this evolution extends over ~30 km (Fig. 3A). Consequently, stations located near the transition may record intermediate particle-motion patterns. In models with short transient speed changes, a transient subshear episode causes a more pronounced change than a transient supershear episode (Fig. S8). These results highlight a limitation of this method in identifying short-lived supershear episodes, which, to the best of our knowledge, is a common challenge for all existing approaches.

Large strike-slip earthquakes may rupture multiple segments with complex fault geometry. We synthesize particle motions for bends of 10°–30° (Fig. S9). The results show that bends mainly affect stations along the segment beyond the bend. A bend of 10° or less has a negligible effect, whereas a 30° bend causes a subshear rupture to produce enhanced early FP motion, i.e., a spurious supershear-like signature, at stations on the inner side within ~10 km of the bend. We also test the sensitivity of particle motion to the dip angle and rake angle by considering a dip angle of 75° and oblique slip with rake angles of 160° (oblique-reverse) and -160° (oblique-normal). The two parameters can modulate the motion to some degree but the overall motion trend is preserved (Fig. S10).

We consider the particle-motion trend unreliable for discriminating rupture speed when it is strongly influenced by the initial rupture expansion, rupture termination, or fault bends. Based on the synthetic tests, particular attention should be paid to stations within ~10 km of the hypocenter along strike, within ~5 km of the rupture termination or beyond the rupture extent, and on the inner side within ~10 km of a fault bend of 30° or larger.

**Near-fault particle motions in M7+ strike-slip earthquakes**

In this section, we investigate near-fault strong-motion data (within 3 km of the fault) for M7+ predominantly strike-slip earthquakes, comprising 20 station records from 8 events (Fig. 1 and Table S2), 13 of which are from the 2023 Mw 7.8 Kahramanmaraş earthquake. Note that the SKR record of the 1999 Izmit earthquake is excluded due to the absence of the NS component *(5)*. We correct for baseline shifts (Fig. S11) and verify that the choice of correction does not substantially alter the particle-motion trend (Fig. S12). We then estimate the local strikes (Table S3) using surface rupture traces and rotate data into FP and FN directions (Fig. S13). All data processing details can be found in Materials and Methods. Velocity pulses and displacement trajectories are plotted (Figs. S14-S21).

No filtering is required in the particle motion analysis. Previous studies have suggested that the radiation in near-fault data may lose coherence at high frequencies owing to isotropic source or other structural effects *(24, 28)*, which may bias the particle-motion pattern. To assess the robustness, we check waveforms and particle motions filtered below 0.5 Hz. The trends of the particle motions are preserved after filtering with only some loss of detail (Figs. S14-S21).

Before inferring the rupture speeds, we apply the criteria derived in the previous section. Stations CLC (Ridgecrest) and KMMH16 (Kumamoto) are located within 10 km of the hypocenters (Fig. S13) and show FP-dominant motion, as predicted for the initial rupture stage (Figs. 3C, S18, and S19). Stations LUC (Landers) and 3139 (Kahramanmaraş) are located on the inner side of significant bends (~30°) (Fig. S13). In particular, station LUC shows a linear motion trend predicted for a subshear rupture passing through a bend (Fig. S14), whereas 3139 does not show the typical subshear or supershear features predicted near bends (Fig. S20). All stations except ROLC are at least 5 km from the ends of the surface rupture traces (Fig. S13), and none shows the oblique displacements predicted near rupture terminations. Although ROLC is close to the end of the surface rupture, its predominantly FP static displacement (Fig. S16) indicates that the rupture extended farther east at depth (*29*). In addition, station KEKS recorded at least three pulses in both the FP and FN components (Fig. S17), which are likely associated with multiple slip episodes on a complex fracture network (*30*). We therefore mark KMMH16, CLC, LUC, 3139, and KEKS as inconclusive in the following rupture speed inferences.

We then identify particle-motion features among the remaining stations. Four stations, NPW (Myanmar), PS10 (Denali), 4615 (Kahramanmaraş), and NAR (Kahramanmaraş), exhibit typical supershear FP-dominant motion (Fig. 1B). Some stations along the southern Amanos segment display typical subshear FN-FP-FN motion patterns (Fig. 1C). Notably, station 3143 shows a bimodal motion pattern, as predicted for a supershear transition stage (Fig. 3D) in which the particle motion separates into an initial supershear FP-dominant motion followed by an FN-dominant motion with the trailing Rayleigh pulse. Stations 3137 and ROLC also show a similar trend with an FP motion followed by an FN pulse (Figs. S16 and S20).

### Particle motion shape metrics for rupture speed determination

The analyses above demonstrate that the shape of particle motion contains first-order information regarding rupture speed. To quantify this information, we define two shape metrics: the normalized particle-motion area and the particle-motion anisotropy (Fig. 4; see details in Materials and Methods). Subshear particle motions are characterized by a large area and low anisotropy, whereas supershear particle motions feature a small motion area and high anisotropy. We calibrate the metrics using kinematic synthetics with off-fault distances of 1-3 km and rupture lengths of 40-120 km. Synthetics with different rupture speeds occupy distinct regions of the metric space, which we separate into supershear ($Vr/Vs>\sqrt{2}$), intermediate ($1<Vr/Vs<\sqrt{2}$), and subshear ($Vr/Vs<1$) regimes (Fig. 4A).

We then apply the metrics to the observations (Fig. 4, B and C). Stations NPW, PS10, NAR, 4615, and 3137 fall in the supershear regime; stations 3138, 3143, 3144, and ROLC fall in the intermediate speed regime; and stations 2708, 2712, 2718, 3142, 3145, and 4616 fall in the subshear regime. This distribution is robust considering the uncertainties of local strikes and the choice of the data window length (Figs. S22 and S23; see details in Materials and Methods). Notably, some stations excluded by our criteria fall in the supershear regime due to their FP-dominant particle motions, which confirms the necessity of these criteria.

We finally compare our results with rupture speeds reported in previous studies (Fig. 4B; Table S4; see details in Materials and Methods). Previous studies on the Myanmar earthquake broadly agree on a sustained supershear rupture near station NPW (*15–17*, *23*, *31–33*). For the Denali earthquake, previous studies commonly suggest that the supershear rupture initiated ~20 km before reaching station PS10 (*14*, *19*, *22*). The rupture speed near stations 4615 and NAR during the 2023 Mw 7.8 Kahramanmaraş earthquake is debated. Studies based on velocity pulses or wave polarization at nearby stations suggest supershear speeds (*21*, *34*), whereas some kinematic models with subshear speeds can also fit most of the data (*35*, *36*). Along the Amanos segment, previous studies generally agree on an overall subshear speed with transient supershear episodes (*13*, *35–39*). In particular, the timings of velocity pulses suggest an apparent supershear propagating speed along the segment with stations 3143, 3138, 3144, and 3137 (*13*). For the Darfield earthquake, the total rupture duration and rupture extent in their preferred kinematic source model (*40*, *41*) imply a subshear average rupture speed, whereas the local rupture speed near station ROLC was not constrained in the past. Through comparing the reported rupture speeds and their distribution in the shape metric domain, our results are generally consistent with previous studies (Fig. 4B).

The particle-motion analysis and the shape metrics are intended to provide a quick and independent indication of rupture speed rather than a definitive estimate. The comparison with previous studies supports the effectiveness of this method in identifying potential supershear ruptures. We note, however, that our synthetic models are still simplified compared with real rupture processes. A more precise estimate would require accounting for additional complexities, such as the continuity of the rupture front, propagation path, local velocity structure, and so on. In addition, the current dataset is limited, with 13 of the 20 records from a single earthquake. Station-level statistics should be interpreted with caution. With the rapidly growing strong-motion network, we anticipate to have more near-fault data that can be applied with our new method.

## Conclusion

In this study, we characterize near-fault displacement particle motions associated with subshear and supershear rupture speeds using kinematic and dynamic rupture models and near-fault records of M7+ strike-slip earthquakes. We demonstrate how rupture heterogeneities can influence the motion trend and identify features associated with initial rupture expansion and supershear transition in real records. Moreover, we propose shape metrics to discriminate rupture speeds from the motion trajectories and validate them against previous studies. Our study highlights the application of near-fault data for providing a quick and independent indication of local rupture propagation speed.

**Acknowledgements**

The authors sincerely thank Nadia Lapusta, Thomas Heaton, Zhongwen Zhan, Ralph J. Archuleta, Yehuda Ben-Zion, Bill Ellsworth, Zhigang Peng, Yoshihiro Kaneko, and Paul Martin Mai for their valuable discussions in Caltech (April 2025) and 2025 SSA Annual Meeting.

**Funding:**

HY and SY gratefully acknowledge the National Key R&D Program of China project (2023YFF0803202), HKSAR RGC projects (14307226, 14308523) and 1+1+1 CUHK-CUHK(SZ)-GDSTC Joint Collaboration Fund (2025A0505000082, 2025A0505000079). SY, HA and HSB gratefully acknowledge funding provided by the ANR project SMEC (ANR-23-CE49-0004). HSB and SY also acknowledge partial support provided by the European Research Council grant PERSISMO (grant 865411).

**Author Contributions:**

Suli Yao contributed to the conceptualization of the research problem, data processing and model result analysis, figure preparation, and manuscript writing and revision. Hongfeng Yang contributed to the conceptualization of the research, project supervision, funding acquisition, and manuscript writing and revision. Harsha S. Bhat contributed to the methodology development, discussion and interpretation of results, and funding acquisition. Hideo Aochi contributed to discussion and interpretation of results, manuscript revision, and funding acquisition.

**Data Availability:** The source of the raw data is listed in table S2. All other data needed to evaluate the conclusions in this paper can be found at Zenodo: (https://doi.org/10.5281/zenodo.23807891).

[This dataset will be publicly available via the above DOI once this study gets published. Now the dataset can be accessed via the following private link: https://zenodo.org/records/23807891?preview=1&token=eyJhbGciOiJIUzUxMiJ9.eyJpZCI6IjUzNmFmMWJmLWJjNDMtNGUyYy1iNGEwLTFmNDVjYjhlNTgzYSIsI

mRhdGEiOnt9LCJyYW5kb20iOiI0NmM1ZDBkZDVlNWQxYmQ1MmNkOTA0YjVhNzlkMTM3NyJ9.hQI-bgJyz1wUFmyQ74gfXa7xTZ8BXWuiSjh0uz5UHSkLed88s1MbT0OEdAOniY4GuAY9cu2VoErM5Yr1QdaRAg.]

**Competing interests:** ALL authors declare no competing interests.

**Supplementary materials:**
Materials and Methods
Table: S1-S4
Figure: S1-S23

## Figures

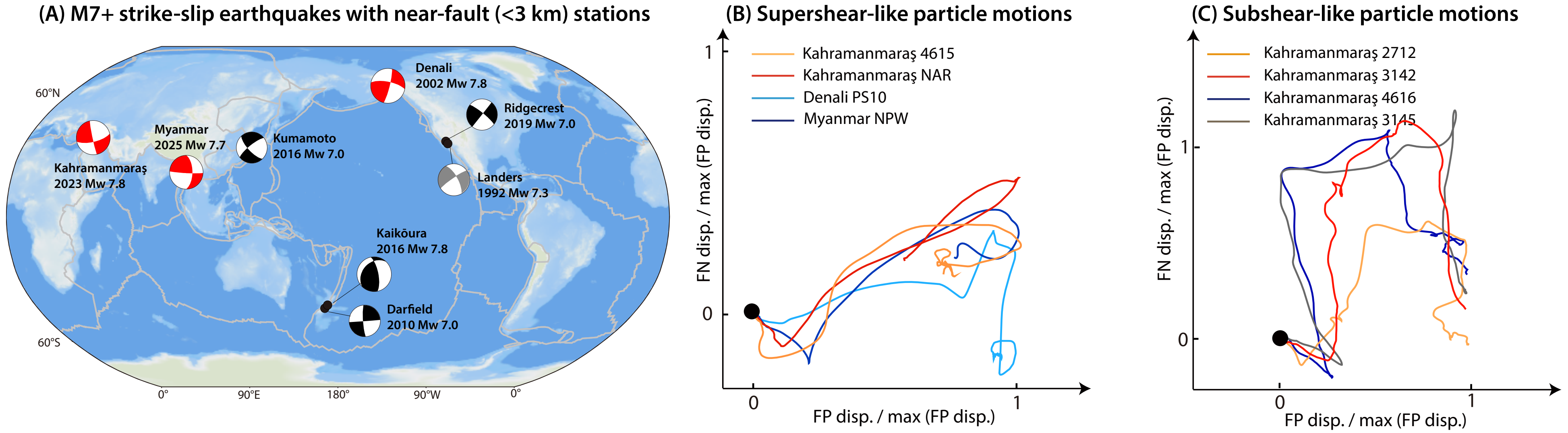

**Figure 1. Near-fault particle motions in M7+ strike-slip earthquakes.** **(A)** Map for M7+ strike-slip earthquakes with near-fault strong-motion records (within 3 km). The focal mechanisms are downloaded from the GCMT catalog (https://www.globalcmt.org) and are colored according to previous studies: red indicates earthquakes with reported supershear rupture, gray indicates the debated case, and black indicates earthquakes with no supershear reports. **(B)** Identified typical supershear particle motions. **(C)** Identified typical subshear particle motions.

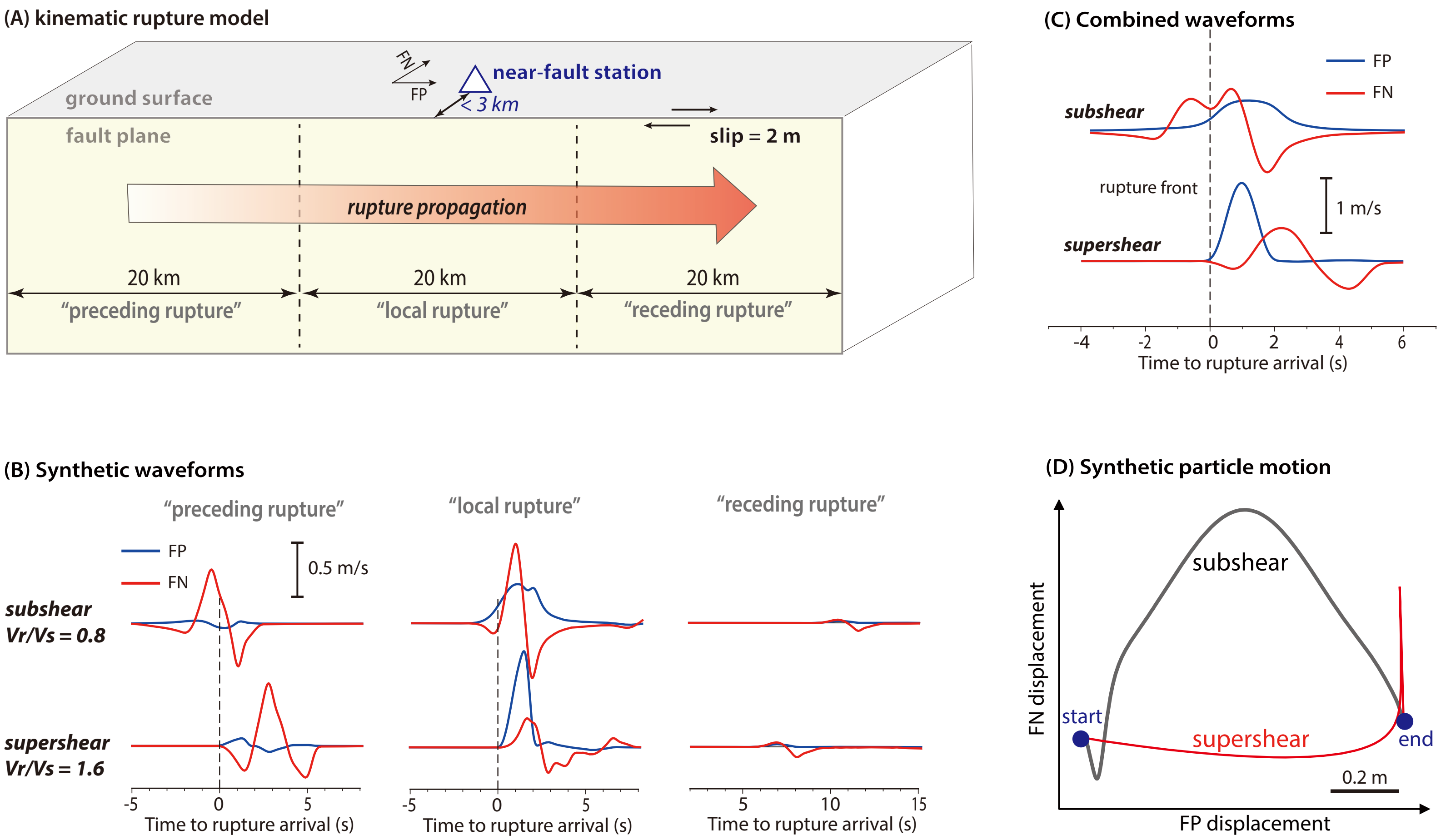

**Figure 2. Synthetic near-fault particle motions in a strike-slip earthquake. (A)** A kinematic strike-slip rupture model and a near-fault station. The total rupture length and the depth extent are 60 km and 10 km, respectively. The station is located at the along-strike center with an off-fault distance of 1 km. **(B)** The kinematic synthetics of fault-normal (FN) and fault-parallel (FP) velocity pulses caused by the preceding, local, and receding rupture for subshear and supershear rupture speeds. The dashed lines mark the local rupture time at the station. **(C)** The synthetic FP and FN velocity waveforms for subshear and supershear rupture speeds. **(D)** Synthetic displacement particle motions in subshear (dark gray) and supershear (red) ruptures at the station in the kinematic model.

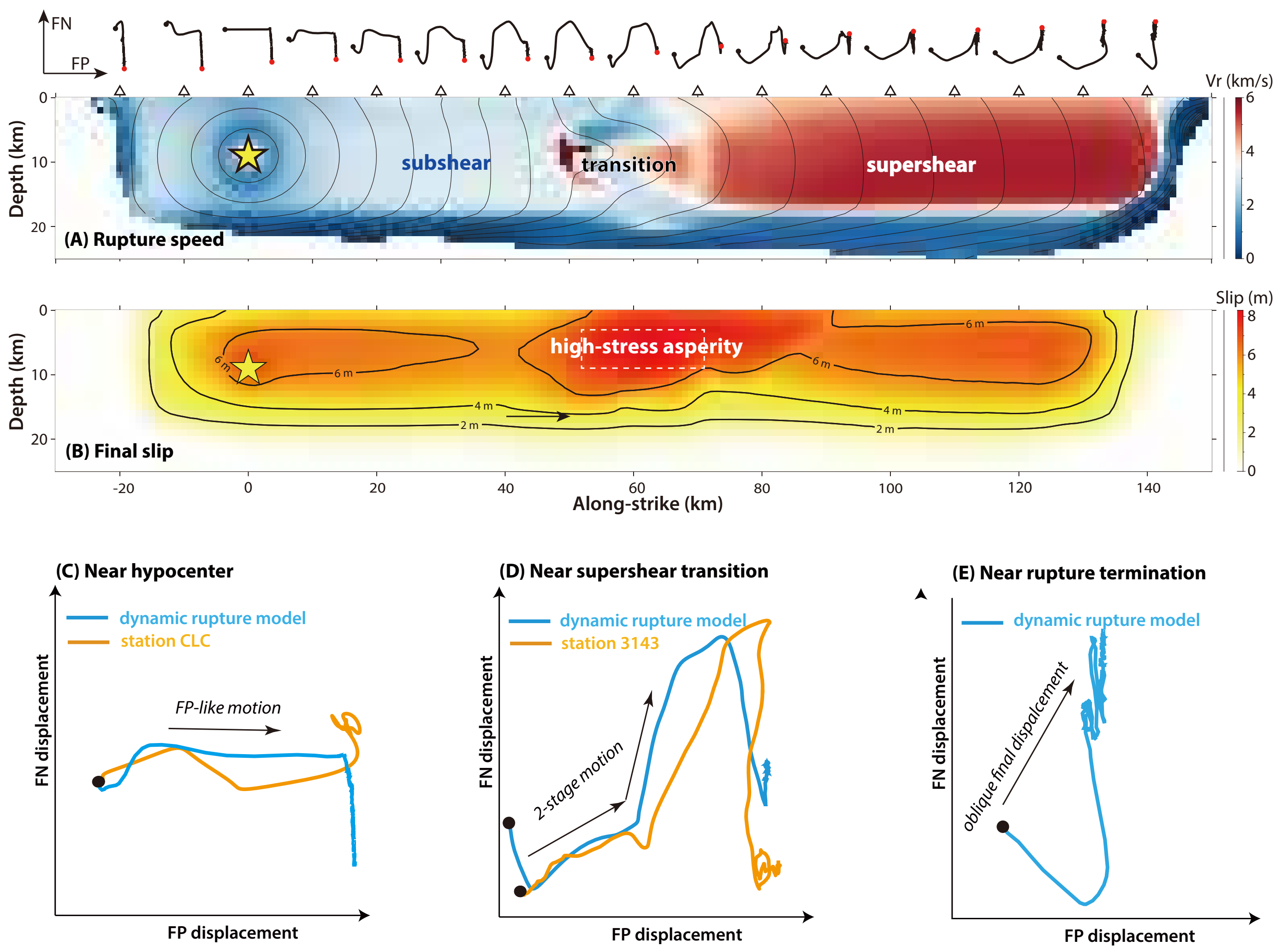

**Figure 3. Particle motions in a dynamic rupture model with a subshear-supershear transition. (A)** The rupture speed and rupture front contours (every 2 seconds). The rupture starts at a depth of 9 km and then propagates unilaterally at a subshear speed. Then the rupture transitions into supershear at a high-stress asperity. Particle motions at an off-fault distance of 1 km along the fault are shown at the top. **(B)** The final slip distribution in the dynamic model. The dashed rectangle marks the location of the high-stress asperity. **(C)** The particle motion near the hypocenter in the dynamic rupture model and at station CLC. **(D)** The particle motion near the subshear-supershear transition in the dynamic rupture model and at station 3143. **(E)** The particle motion near the rupture termination in the dynamic rupture model. Each particle motion is uniformly scaled by its maximum FP displacement for comparison, preserving its shape and aspect ratio.

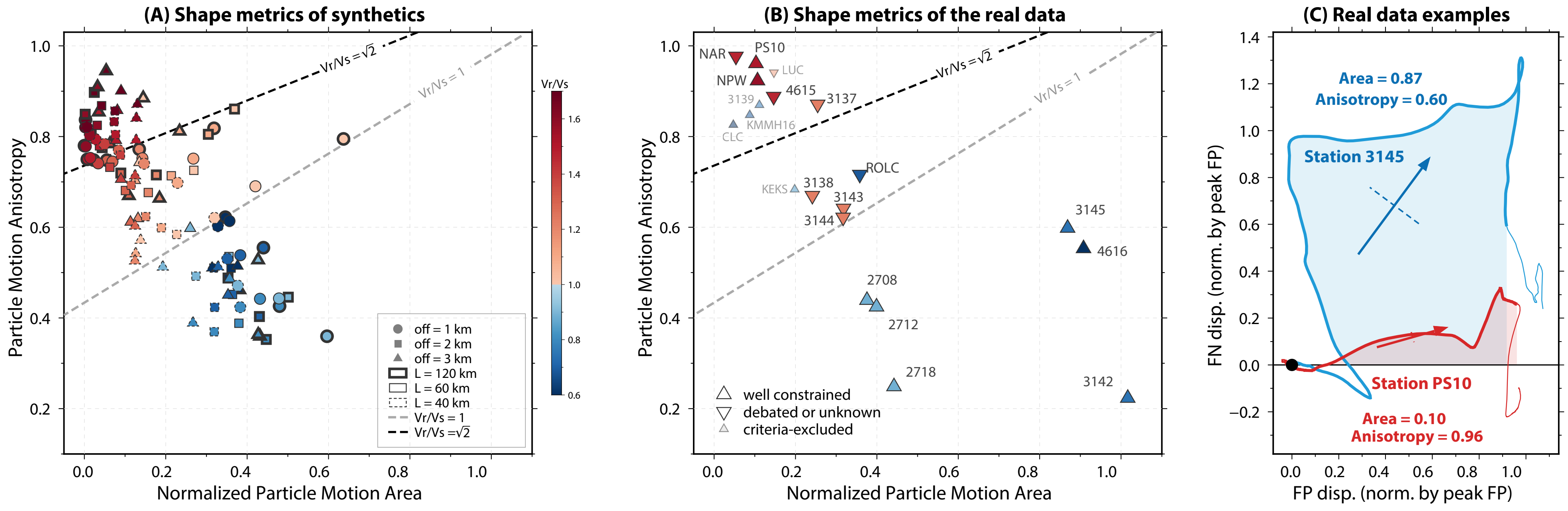

**Figure 4. Particle motion shape metrics. (A)** The normalized particle motion area and anisotropy of synthetic particle motions in kinematic models. Synthetics at different off-fault distances and rupture lengths are marked by different symbols. The symbols are colored according to the ratio of rupture speed to the shear wave velocity ($V_r/V_s$). The dashed lines mark the boundaries between regimes of subshear, intermediate, and supershear rupture speeds. **(B)** The shape metrics of observed near-fault station records. All stations are colored according to the rupture speeds reported in previous studies. Stations excluded by the criteria (near hypocenter or on the inner side of a significant fault bend) are marked as small triangles. Stations with debated or unknown local rupture speeds are marked as inverted triangles. **(C)** The particle motions and the shape metrics at stations 3145 and PS10, as the typical subshear and supershear cases, respectively. The thick lines mark the particle motions within a time window from −5 to 5 seconds relative to the peak horizontal velocity. The light blue and light red shaded areas represent the particle motion area. Arrows and dashed lines indicate the major and minor axis orientations and relative amplitudes from anisotropy analysis.

Supplementary Materials for

# Supershear Rupture Indicator in Near-fault Particle Motion

Suli Yao[1,2], Hongfeng Yang[1,3,4,5*], Harsha S. Bhat[2], and Hideo Aochi[2,6]

[1]Department of Earth and Environmental Sciences, The Chinese University of Hong Kong, Hong Kong, China

[2]Laboratoire de Géologie, École Normale Supérieure, CNRS-UMR 8538, PSL Research University, Paris, France

[3]Shenzhen Research Institute, The Chinese University of Hong Kong, Shenzhen, Guangdong, China

[4]Institute of Environment, Energy and Sustainability, The Chinese University of Hong Kong, Hong Kong, China

[5]Institute of Space and Earth Information Science, The Chinese University of Hong Kong, Hong Kong, China

[6]Bureau de Recherches Géologiques et Minières (BRGM), 45100 Orléans, France

* Corresponding author: hyang@cuhk.edu.hk

## The PDF file includes:

Materials and Methods

Tables S1-S4

Figures S1-S23

# Materials and Methods

## 1, Kinematic modeling of near-fault ground motion

We model the near-fault ground motion in strike-slip ruptures on a vertical planar fault using a kinematic framework. The ruptures extend from the ground surface to a depth of 10 km and have varying along-strike extents. We divide the fault into subfaults with a spacing of 0.5 km and employ the PyGRT tool (*1*) to compute the Green's functions from each subfault to the near-fault station. A homogeneous half-space velocity structure ($V_P$ = 6.0 km/s, $V_s$ = 3.5 km/s) is adopted. The sampling rate is 50 Hz. Synthetic stations are positioned along the fault with varying off-fault distances. In the model in Figure 2, models with bends (Fig. S9), resolution tests (Fig. S3), models with different slip durations (Fig. S7), and tests of rake and dip angles (Fig. S10), the rupture is assumed to propagate unilaterally with a vertical rupture front at a constant rupture propagation speed, thereby neglecting the rupture initiation process. In other models (Figs. S5, S6, and S8), the rupture is assumed to propagate from the hypocenter at a constant rupture speed with a curved rupture front. At each subfault, the source time function is assumed to follow $\delta_{final} * sin^2(\frac{t}{T}\pi)$ $(0 \leq t \leq T)$, where the slip duration T is 2 s and the final slip $\delta_{final}$ is 2 m. We test the influence of the slip pulse duration by varying it from 1 to 3 seconds and find it to be minor (Fig. S7).

To test the sensitivity of the particle motion to the rupture speed, we model the particle motions at different off-fault distances (1-20 km) with different rupture lengths (Figs. S3). We place the stations at the along-strike center of the fault. The rupture propagates unilaterally with a vertical rupture front. Given a total rupture length of 60 km (i.e. extending 30 km on each side of the station), the features distinguishing subshear and supershear motions are preserved at off-fault distances up to ~8 km (Fig. S3). As the rupture length decreases to 20 km, the particle motion trends are nearly identical in subshear and supershear ruptures at an off-fault distance of $\geq$5 km, indicating a degraded discriminability. We measure the shape metrics for all synthetics and find that stations within 3 km can stably differentiate subshear and supershear for a rupture extends 20 km from the station (total length L=40 km) (Fig. S4). This test provides a rough estimate of the method resolution.

We construct heterogeneous slip distributions for kinematic models (Fig. S6). We first define a deterministic background of depth-dependent slip distribution with the max slip occurring at the depth of 5-8 km and a shallow slip deficit of 30%. Based on this, we then add stochastic heterogeneities. The stochastic texture is generated from a von Kármán random field (*2*) (correlation lengths of 3 km along strike and 2 km along dip, Hurst exponent of 0.6) and applied as a multiplicative perturbation. The resulting slip is allowed to vary between 0.15 and 1.63 times the mean background slip at the depth. We further test another model with a large slip asperity. The background setting of overall depth-dependent slip distribution is kept. We add a Gaussian asperity centered at 3.5 km depth, with an along-strike width of 18 km and a depth extent of 4 km. The average slip inside the asperity is twice the average slip outside the asperity. The slip on the fault varies between 0.31 to 2.45 times the mean slip.

## 2, Dynamic rupture simulation with supershear transition

We perform a 3-D dynamic rupture simulation on a vertical planar strike-slip fault embedded in a cubic elastic domain extending 180 km along strike, 80 km along the strike-normal direction, and 50 km in depth. The rupture extends along strike for 140 km with dominantly unilateral rupture propagation. The grid size on the fault is 150 m. The time step of the simulation is 0.005 s. We adopt the 1-D velocity model for the 2023 Kahramanmaraş Mw 7.8 earthquake region to prescribe the elastic material properties (*3*) (Table S1).

We adopt a linear slip-weakening friction law (*4*) (Eqs. 1 and 2), in which the frictional strength ($\tau_f$) decreases linearly with fault slip (δ) from the yield stress ($\tau_s$) to the dynamic stress ($\tau_d$) when the slip accumulates to the critical weakening distance ($D_c$). We set typical values of 0.7 and 0.4 for the static ($\mu_s$) and dynamic ($\mu_d$) friction coefficients inside the seismogenic zone. The seismogenic zone extends from 2 to 16 km in depth. Outside the seismogenic zone, the fault gradually transitions into strengthening with a dynamic friction coefficient (i.e., 0.9) higher than the static coefficient. The widths of the shallow and deep transition zones are 1 km and 4 km, respectively. The effective normal stress ($\sigma_n$) and initial stress ($\tau_i$) are 50 MPa and 28 MPa below 3 km, respectively, and both gradually decrease to 0 at the free surface. We set the cohesion strength ($C$) to 3 MPa at the free surface. The cohesion strength decreases linearly with depth from 3 MPa at the free surface to 0 at a depth of 3 km. $D_c$ is 1 m.

$$\tau_f(\delta) = \begin{cases} \tau_s - \frac{\delta}{D_c} \times (\tau_s - \tau_d) + C & \delta < D_c \\ \tau_d + C & \delta \geq D_c \end{cases} \tag{1}$$

$$\tau_d = \mu_d \times \sigma_n \quad ; \; \tau_s = \mu_s \times \sigma_n \tag{2}$$

We nucleate the rupture at a depth of 9 km by increasing the initial shear stress to 0.1 MPa above the yield stress inside a circular nucleation zone with a radius of 3 km. The rupture first expands radially and then propagates laterally. The rupture propagates along strike at a subshear rupture speed for 50 km (Fig. 3). Then a supershear transition is stimulated by increasing the initial stress by 4 MPa between x=50 km and x=70 km at the depth of 6-16 km (Figs. S1 and 3B). The rupture then propagates at a sustained supershear rupture speed. The final total along-strike rupture length is ~140 km extending from the ground surface to ~20 km deep (Fig. 3B). In addition, we output the synthetic near-fault ground motions at different along-strike locations for analysis to capture characteristics of particle motions during the initial rupture expansion, subshear rupture, supershear transition, and sustained supershear rupture (Fig. 3A). We use a finite-element package, PyLith (version 2.2.2) (*5*, *6*), to run the simulation. All model inputs are shown in Figure S1.

To check the resolution, we examine the number of grid points within the cohesive zone during the dynamic rupture. As shown in Figure S2, we take two snapshots of slip at the subshear and supershear stages. We measure the cohesive zone width at the rupture front as the distance between the points with slip of 0.01 m and 1 m (i.e., $D_c$), where the weakening occurs. The cohesive zone is at least 0.75 km wide in the subshear stage and at least 1.02 km wide in the supershear stage, corresponding to at least 5 elements ($\Delta x = 150$ m).

**3, Strong-motion data compilation**

We collect strong-motion data recorded within 3 km of the fault trace for eight M7+ predominantly strike-slip earthquakes reported in previous literature. Kaikōura earthquake has remarkable reverse component and whether it mainly occurred on some shallow crustal faults or the deep subduction interface is debated (*7*). We check the station records and find the vertical displacement at KEKS is ~1.5 m, much smaller than the horizontal displacement of ~7 meters. Therefore, we believe the rupture near this station is dominantly strike-slip. The data sources are listed in Table S2. While the 2023 Mw 7.8 Kahramanmaraş earthquake has records from 13 stations, each of the other 7 earthquakes only have a single station record. We estimate the off-fault distance to the mapped surface rupture traces (Fig. S13). Then we estimate the local strikes by fitting the surface rupture traces using segments of ±5 km or ±10 km of each station. The differences in strike between ±5 km and ±10 km segments are up to ~7°. All information and the station locations are listed in Table S3. The influence of the local strike on results will be discussed later.

Inspection of the data reveals baseline shifts during violent shaking at some stations. Baseline shifts typically manifest as linear or curved trends superimposed on the velocity waveforms. We follow the correction method introduced in (*8*) to remove the drifts (Fig. S11). Then we integrate the velocity waveforms to displacement series for particle motion analysis. We check the corrected waveforms and compare the static displacements with InSAR-based deformation fields (*9–13*), without forcing an exact match because the InSAR data also contain processing uncertainties and possibly postseismic deformation. The corrected displacements are broadly consistent with the geodetic observations (Fig. S11).

Because the large pulses usually dominate over the long-period baseline trends, the motion during the large pulses is largely unaffected by the correction. To verify this, we compare the trajectories within ±10 s of the peak horizontal velocity from the minimally processed and the fully corrected data (fig. S12). For the minimally processed data, we only remove the linear trend in the velocity estimated from the data before the window of large pulses. The trajectories from the two datasets are very similar at all stations except for the late-stage motions at 2708 and 3138. Overall, the particle-motion trends are robust.

The velocity and displacement series are subsequently rotated into fault-parallel (FP) and fault-normal (FN) components according to the local strike (±5 km). Here we assume that the FP direction aligns with the local slip direction. The positive FN direction is defined as the S-wave particle motion direction of the preceding rupture. The velocity waveforms and the particle motion after rotation at each station are plotted in Figures S14-S21. The FP and FN directions at each station are written on panels.

**4, Particle motion shape metrics**

To quantify the differences in motion shape, we test the following two shape metrics. The first metric is the normalized particle motion area. To calculate it, we first normalize the particle motion with the difference between the maximum and minimum FP displacement inside the window ($\max(u_{FP}) - \min(u_{FP})$). In this case, only the shape matters instead of the absolute amplitude. Then the motion area is defined as the area enclosed by the normalized trajectory and the line of FN = 0. This metric is sensitive to the shape of the particle motion and the relative amplitude of FN motion to the FP motion.

Particle motions in subshear ruptures exhibit a larger motion area due to their FN-FP-FN shape and relatively large FN motion, whereas supershear cases tend to feature a smaller area with the dominant FP motion (Fig. 4).

Another metric is the particle motion anisotropy. Let the demeaned displacement components be:

$$u_1 = u_{FP} - \langle u_{FP} \rangle, \qquad u_2 = u_{FN} - \langle u_{FN} \rangle \tag{3}$$

where $\langle \cdot \rangle$ denotes the time average over the window. We form the 2×2 covariance (second-moment) matrix of the particle motion:

$$C = \begin{pmatrix} \langle u_1^2 \rangle & \langle u_1 u_2 \rangle \\ \langle u_1 u_2 \rangle & \langle u_2^2 \rangle \end{pmatrix} \tag{4}$$

Its eigenvalues $\lambda_1 \geq \lambda_2 \geq 0$ are the variances of the motion along the two principal axes (the eigenvectors), with $\lambda_1$ along the major axis and $\lambda_2$ along the minor axis. The anisotropy is defined as:

$$A = (\lambda_1 - \lambda_2) / (\lambda_1 + \lambda_2), \qquad A \in [0, 1]. \tag{5}$$

A=0 corresponds to isotropic (circular) particle motion, while A=1 corresponds to rectilinear motion. Anisotropy is thus a measure of how linearly polarized the ground motion is. Particle motions in subshear ruptures exhibit more circle-like motion with lower anisotropy, whereas supershear cases tend to be more rectilinear with higher anisotropy (Fig. 4).

We first calculate the two metrics for synthetics in kinematic models (Fig. 4A). Considering continuous rupture speeds ranging from subshear to supershear, the two metrics show a continuous distribution. Based on the distribution, we define two boundaries corresponding to Vr/Vs=1 and Vr/Vs=$\sqrt{2}$ (Fig. 4A). Then we apply the two metrics to the real near-fault data (Fig. 4B). Considering the pulse durations, all metrics are calculated over the window of −5 to 5 seconds relative to the time of peak horizontal velocity, which captures the major motion trend. The results are stable when the window is extended to ±7 s (Fig. S23). A shorter window increases the anisotropy at some subshear stations, because it is not long enough to cover the whole motion trend and may miss the early or late FN motion. The motion area is less sensitive to the window length, as long as the main FP displacement is included.

The anisotropy metric is rotation-invariant, whereas the area index depends on the strike. We test the influence of the strike with variations of ±10° (Fig. S22). The stations with low area indices remain low. The larger changes at 3145 and 4616 arise because their oblique final displacements make the peak FP displacement, used to normalize the area index, more sensitive to rotation. Both stations nevertheless remain in the subshear regime.

**5, Rupture speeds reported in previous studies**

We validate the results of particle-motion analysis against the rupture speeds reported in previous studies. We mainly consider studies involving the near-fault or local stations as the constraints, as we believe that those models can have better resolution for the local rupture speed. For example, the PS10 station record has been well studied through waveform simulations and a supershear speed has been commonly suggested (*14*, *15*). The same holds for NPW station (*16–18*). However, for the 2023 Mw 7.8 Kahramanmaraş earthquake, despite of near-fault records, the rupture speed is debated, especially on the Narlı splay fault near stations 4615 and NAR. Near-fault data features (*19*) and wave polarization analyses at nearby records (*20*) suggest supershear speeds while some finite-fault inversion models suggest potential subshear speeds (*21*, *22*). Therefore, we mark the two stations as debated supershear cases. Along the southern Amanos fault, previous studies agree on the overall subshear speeds (*20*, *21*, *23*, *24*). The timings of near-fault velocity pulses suggest potential supershear rupture along a flat segment (*25*). Therefore, we mark those stations along that segment as debated supershear.

The rupture speed near the LUC station in the 1992 Landers earthquake is debated. Although previous studies agree that the rupture is overall subshear in this event, some kinematic models suggest local abrupt deceleration and then acceleration at the transition between two segments before reaching the station. The constrained local rupture speeds can vary from 3 km/s to 6 km/s (*26–29*). Therefore, we mark this station as a debated supershear. For the 2016 Mw 7.0 Kumamoto, 2016 Mw 7.8 Kaikōura, and 2019 Mw 7.0 Ridgecrest earthquakes, no supershear speeds have been reported, and we refer to the finite-fault inversion results for the rupture speeds. The rupture speed in the 2010 Mw 7.0 Darfield have not been discussed in detail in previous literature and therefore is marked as unknown. The rupture speeds we adopted in Figure 4B in previous studies and references are listed in Table S4.

## Tables

**Table S1. 1-D velocity model adopted in the dynamic rupture model (*3*).**

| Depth (km) | Vp(km/s) | Vs(km/s) |
|---|---|---|
| 0 | 3.88 | 2.04 |
| 1 | 4.52 | 2.43 |
| 2 | 5.62 | 3.03 |
| 4 | 5.75 | 3.31 |
| 6 | 5.85 | 3.38 |
| 8 | 5.96 | 3.43 |
| 10 | 6.00 | 3.44 |
| 12 | 6.05 | 3.46 |
| 16 | 6.32 | 3.62 |
| 20 | 6.40 | 3.67 |
| 25 | 6.83 | 3.92 |
| 30 | 6.89 | 3.94 |
| 37 | 7.80 | 4.40 |
| 45 | 8.22 | 4.56 |
| 60 | 8.30 | 4.61 |

**Table S2. The information of near-fault data used in this study.**

| Event | Mw (GCMT) | Station | strike/dip/rake (GCMT) | Data source |
|---|---|---|---|---|
| 2025-03-28 Mandalay | 7.7 | NPW | 353/60/175 | GEOFON GFZ https://geofon.gfz.de/ |
| 2023-02-06 Kahramanmaraş | 7.8 | total 13* | 51/70/-4 | AFAD-TADAS https://tadas.afad.gov.tr |
| 2019-07-06 Ridgecrest | 7.0 | CLC | 321/81/180 | CESMD https://www.strongmotioncenter.org/ |
| 2016-11-13 Kaikōura | 7.8 | KEKS | 226/33/141, 350/70/63 | New Zealand Strong motion database https://www.geonet.org.nz/data/supplementary/nzsmdb |
| 2016-04-15 Kumamoto | 7.0 | KMMH16 | 222/77/-161 | Japan NIED KiKnet https://www.kyoshin.bosai.go.jp/en/eqdownload/ |
| 2010-09-03 Darfield | 7.0 | ROLC | 88/89/172 | New Zealand Strong motion database https://www.geonet.org.nz/data/supplementary/nzsmdb |
| 2002-11-03 Denali | 7.8 | PS10 | 296/71/171 | CESMD https://www.strongmotioncenter.org/ |
| | | | | (corrected in (14)) |
| 1992-06-28 Landers | 7.3 | LUC | 341/70/-172 | CESMD https://www.strongmotioncenter.org/ |

***Kahramanmaraş stations:** 2708, 2712, 2718, 3137, 3138, 3139, 3142, 3143, 3144, 3145, 4615, 4616, NAR

**Table S3. The information of all stations.**

| **Station** | **Longitude (°)** | **Latitude (°)** | **Off-fault distance (km)** | **Strike ±5 / ±10 km (°)** | **Rupture trace source** |
|---|---|---|---|---|---|
| LUC | −116.6120 | 34.5680 | 2.0 | 306.5 / 313.5 | (30) |
| PS10 | −145.7627 | 63.4245 | 3.1 | 121.8 / 118.5 | (31) |
| ROLC | 172.3808 | −43.5944 | 2.3 | 265.3 / 260.9 | (32, 33) |
| KMMH16 | 130.8199 | 32.7967 | 0.7 | 43.8 / 43.3 | (34) |
| KEKS | 173.9810 | −41.9560 | 3.0 | 61.0 / 58.4 | (35) |
| CLC | −117.5980 | 35.8160 | 1.9 | 148.1 / 151.7 | (36) |
| NAR | 37.1574 | 37.3919 | 0.7 | 16.4 / 21.4 | (37) |
| 4615 | 37.1380 | 37.3868 | 0.9 | 194.9 / 200.7 | (37) |
| 4616 | 36.8384 | 37.3755 | 2.8 | 212.7 / 216.1 | (37) |
| 2712 | 36.7328 | 37.1840 | 1.2 | 31.8 / 33.2 | (37) |
| 2708 | 36.6484 | 37.0993 | 0.5 | 28.7 / 28.9 | (37) |
| 2718 | 36.6266 | 37.0078 | 1.1 | 12.8 / 17.4 | (37) |
| 3143 | 36.5571 | 36.8489 | 1.6 | 20.8 / 18.1 | (37) |
| 3138 | 36.5112 | 36.8026 | 0.7 | 200.0 / 201.1 | (37) |
| 3144 | 36.4857 | 36.7569 | 0.7 | 203.1 / 204.6 | (37) |
| 3137 | 36.4885 | 36.6929 | 2.5 | 33.5 / 32.9 | (37) |
| 3145 | 36.4064 | 36.6454 | 0.5 | 27.2 / 26.2 | (37) |
| 3139 | 36.4144 | 36.5838 | 2.4 | 9.6 / 14.6 | (37) |
| 3142 | 36.3661 | 36.4980 | 1.3 | 29.2 / 26.4 | (37) |
| NPW | 96.1376 | 19.7785 | 2.5 | 350.6 / 351.4 | (38) |

**Table S4. Reference rupture speed adopted for each station.**

| **Station** | **Earthquake** | **Vr (km/s)** | **Reference** |
|---|---|---|---|
| LUC | 1992 Mw 7.3 Landers | 4.0 | (26) |
| PS10 | 2002 Mw 7.8 Denali | 5.3 | (14) |
| ROLC | 2010 Mw 7.0 Darfield | 2.4 | (39) |
| KMMH16 | 2016 Mw 7.0 Kumamoto | 2.4 | (40) |
| KEKS | 2016 Mw 7.8 Kaikōura | 2.9 | (7) |
| CLC | 2019 Mw 7.0 Ridgecrest | 2.0 | (41) |
| 4615 | 2023 Mw 7.8 Kahramanmaraş | 5.15 | (19) |
| NAR | 2023 Mw 7.8 Kahramanmaraş | 5.15 | (19) |
| 4616 | 2023 Mw 7.8 Kahramanmaraş | 2.0 | (*25*) |
| 2712 | 2023 Mw 7.8 Kahramanmaraş | 3.1 | (*25*) |
| 2708 | 2023 Mw 7.8 Kahramanmaraş | 3.1 | (*25*) |
| 2718 | 2023 Mw 7.8 Kahramanmaraş | 3.1 | (*25*) |
| 3143 | 2023 Mw 7.8 Kahramanmaraş | 4.3 | (*25*) |
| 3138 | 2023 Mw 7.8 Kahramanmaraş | 4.3 | (*25*) |
| 3144 | 2023 Mw 7.8 Kahramanmaraş | 4.3 | (*25*) |
| 3137 | 2023 Mw 7.8 Kahramanmaraş | 4.3 | (*25*) |
| 3145 | 2023 Mw 7.8 Kahramanmaraş | 2.6 | (*25*) |
| 3139 | 2023 Mw 7.8 Kahramanmaraş | 2.6 | (*25*) |
| 3142 | 2023 Mw 7.8 Kahramanmaraş | 2.6 | (*25*) |
| NPW | 2025 Mw 7.7 Mandalay | 5.5 | (*18*) |

## Figures

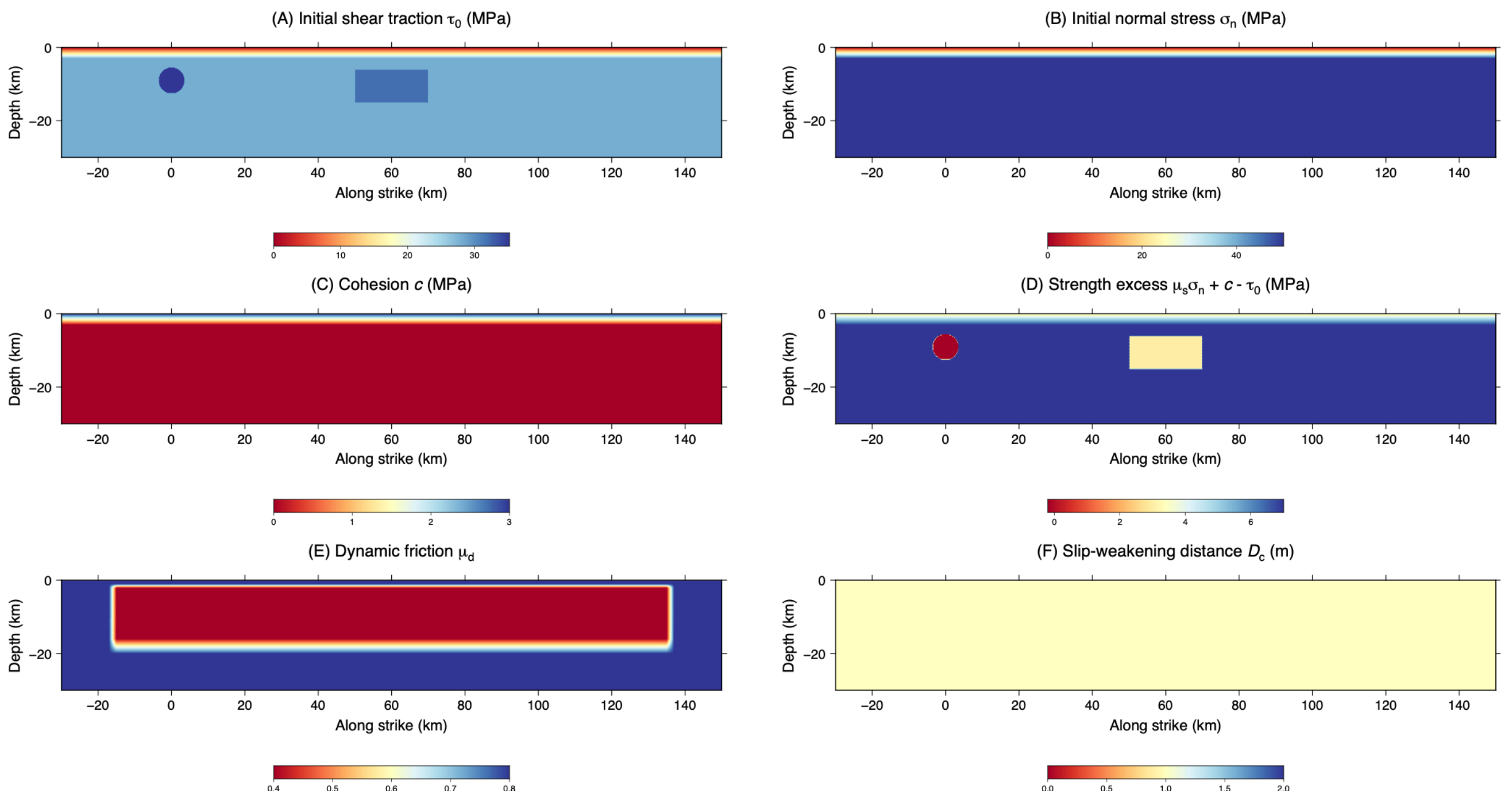

**Figure S1. Inputs of the dynamic rupture model. (A)** Initial shear stress on the fault. **(B)** Effective normal stress. **(C)** Cohesion. **(D)** Strength excess. **(E)** Dynamic friction coefficient. **(F)** Slip-weakening distance.

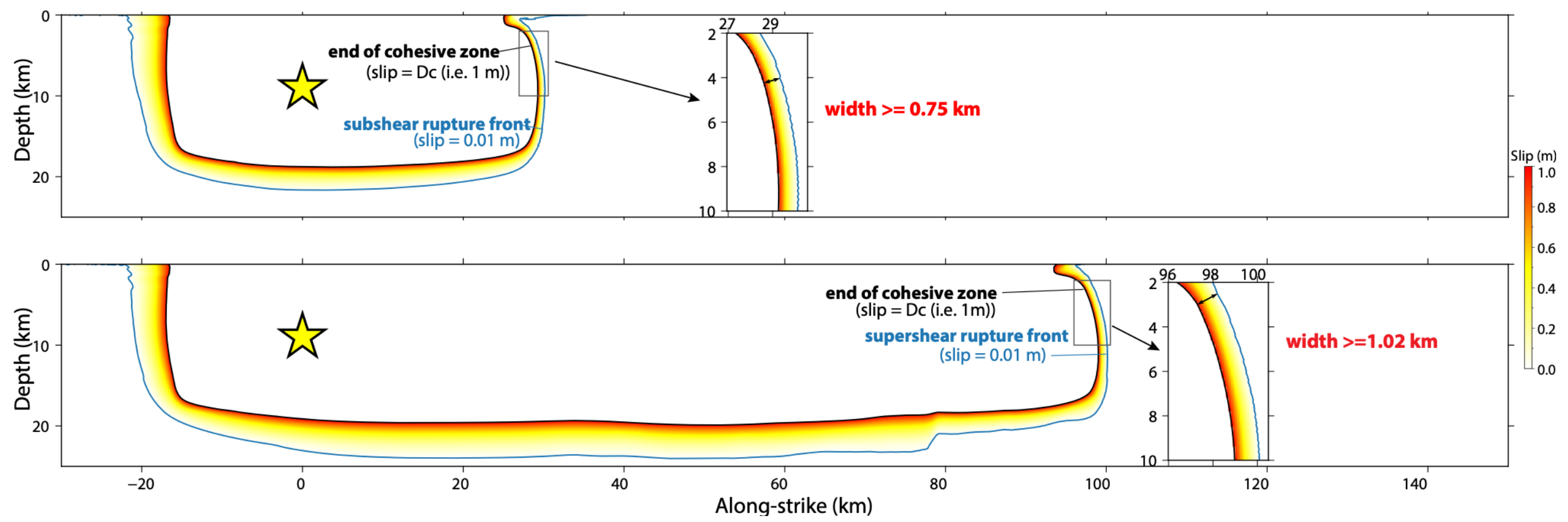

**Figure S2. Snapshots of cohesive zone during subshear (top) and supershear (bottom) rupture in the dynamic rupture models.** The colored zones mark the area with slip between 0.01 m and 1.0 m, where the weakening process is occurring. The minimum width of the zone at the rupture front is estimated and written in red in each panel.

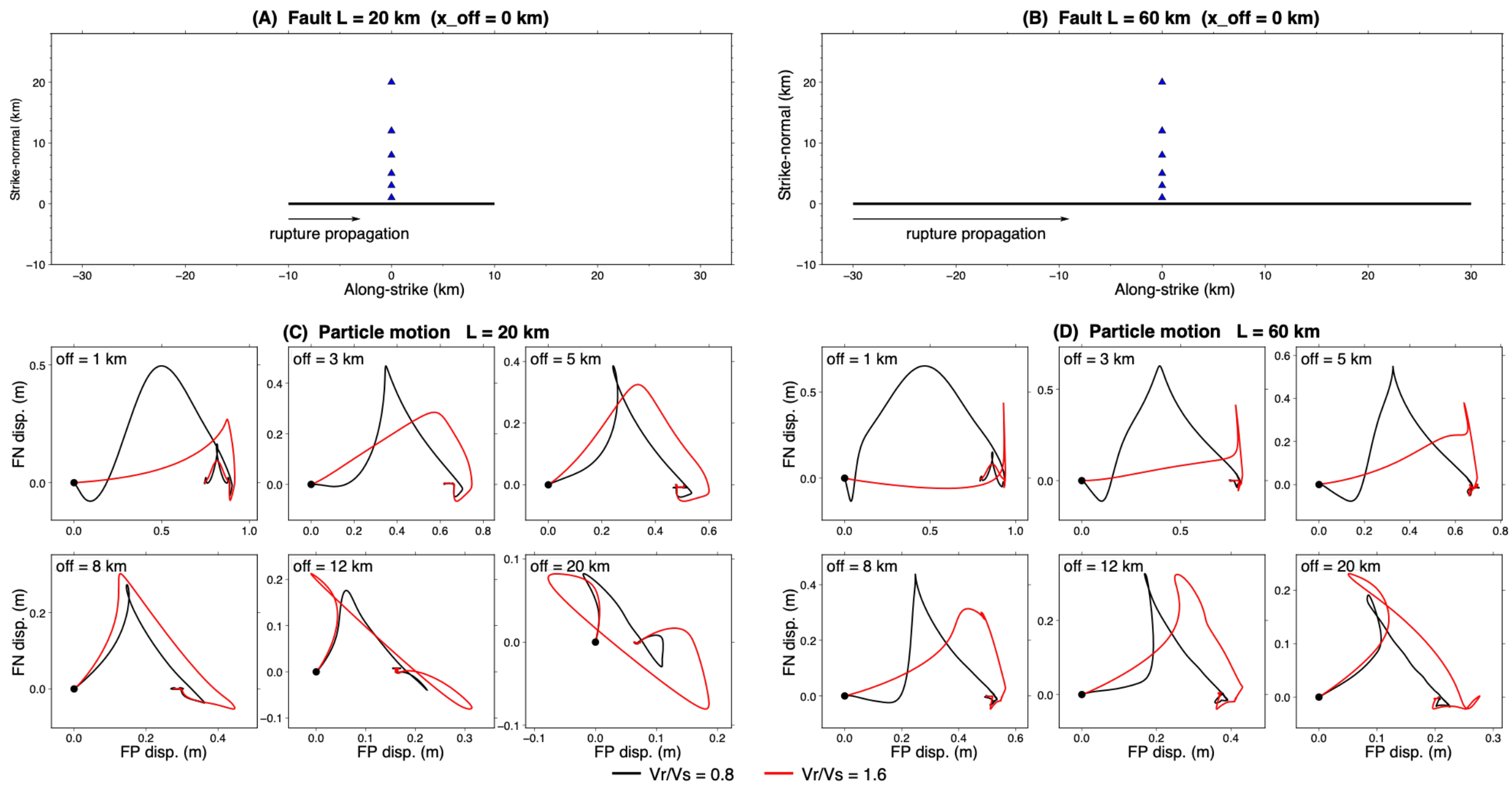

**Figure S3. Particle motions at different off-fault distances and rupture lengths. (A and B)** Fault traces and station locations for rupture lengths of 20 km and 60 km, respectively. The stations are located at the center of the fault with off-fault distances of 1-20 km. **(C and D)** Particle motions at off-fault distances of 1–20 km for subshear (Vr/Vs = 0.8, black) and supershear (Vr/Vs = 1.6, red) rupture, corresponding to (A) and (B), respectively.

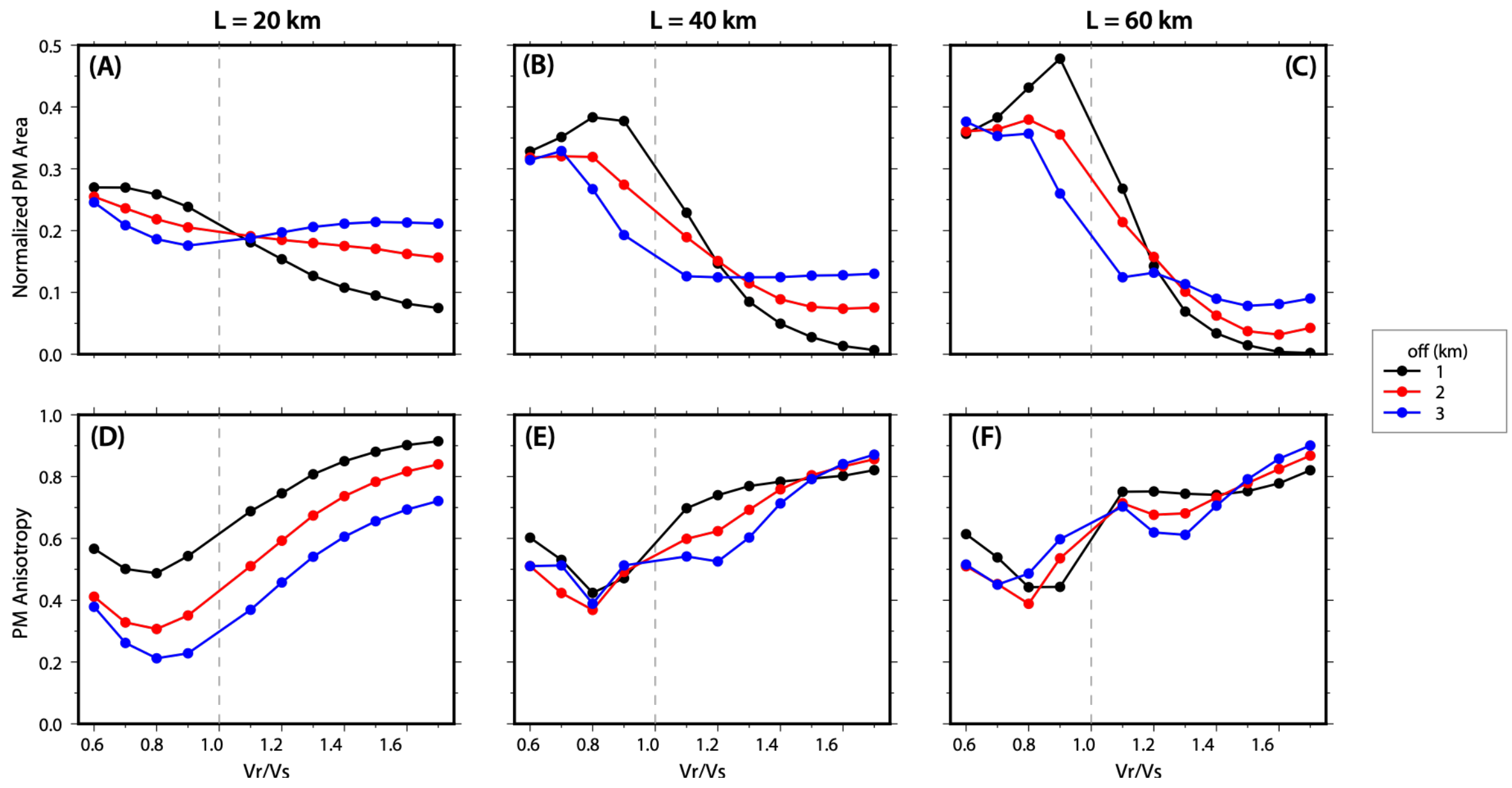

**Figure S4.** Shape metrics of synthetics at different off-fault distances and rupture lengths. **(A-C)** Normalized area index for synthetics with rupture lengths of 20 km, 40 km, and 60 km, respectively. Synthetics at different off-fault distances are marked as different colors. **(D-F)** Same as (A)- (C) but for the motion anisotropy.

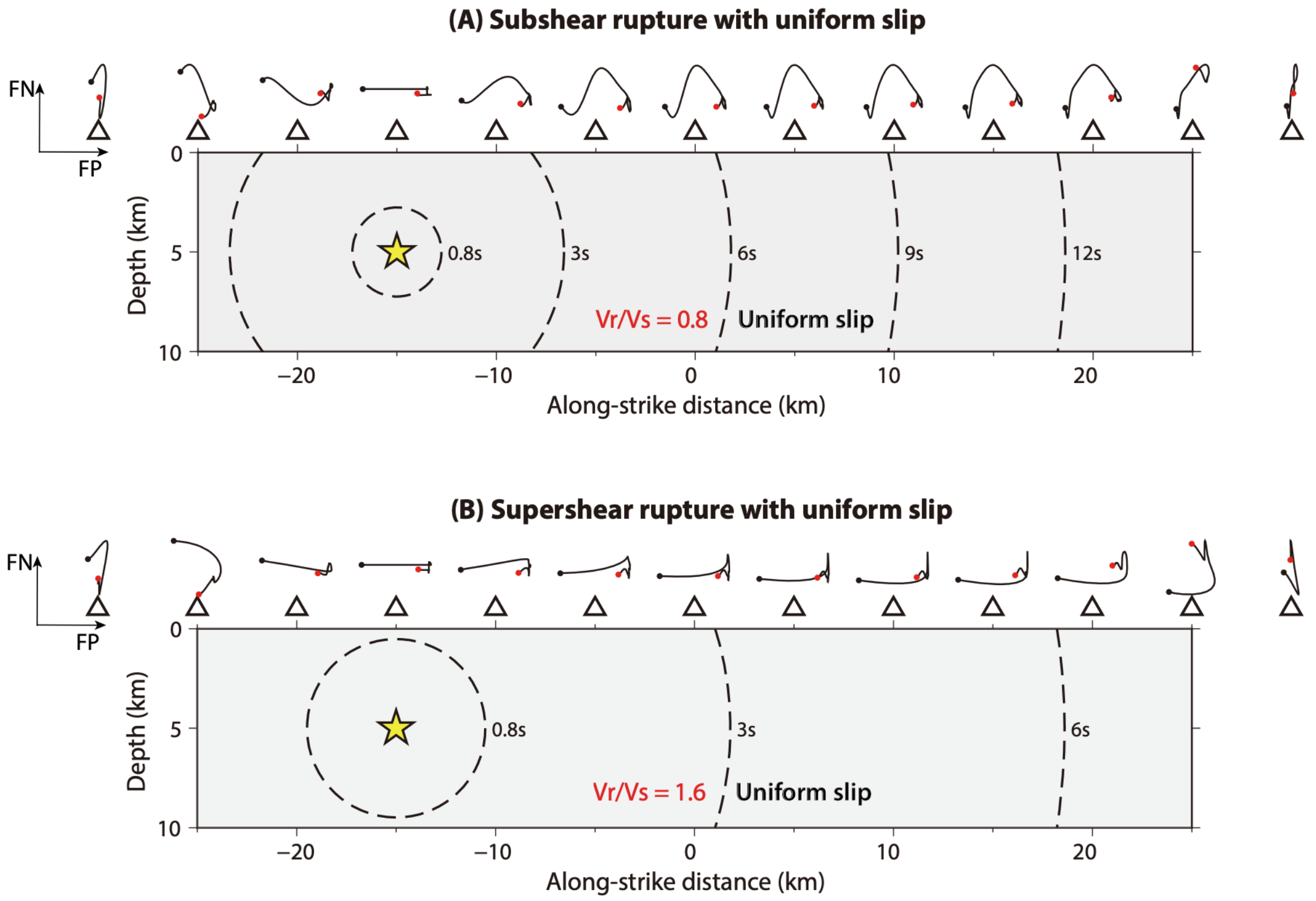

**Figure S5. Kinematic rupture models with uniform slip and the resulting near-fault particle motions. (A)** Subshear rupture (Vr/Vs = 0.8). **(B)** Supershear rupture (Vr/Vs = 1.6). The yellow star denotes the hypocenter, and the dashed lines show the rupture front at 0.8 s and every 3 s. The particle motions at an off-fault distance of 1 km along the fault are plotted at the top of each panel. The black and red dots mark the start and the end of the motion, respectively.

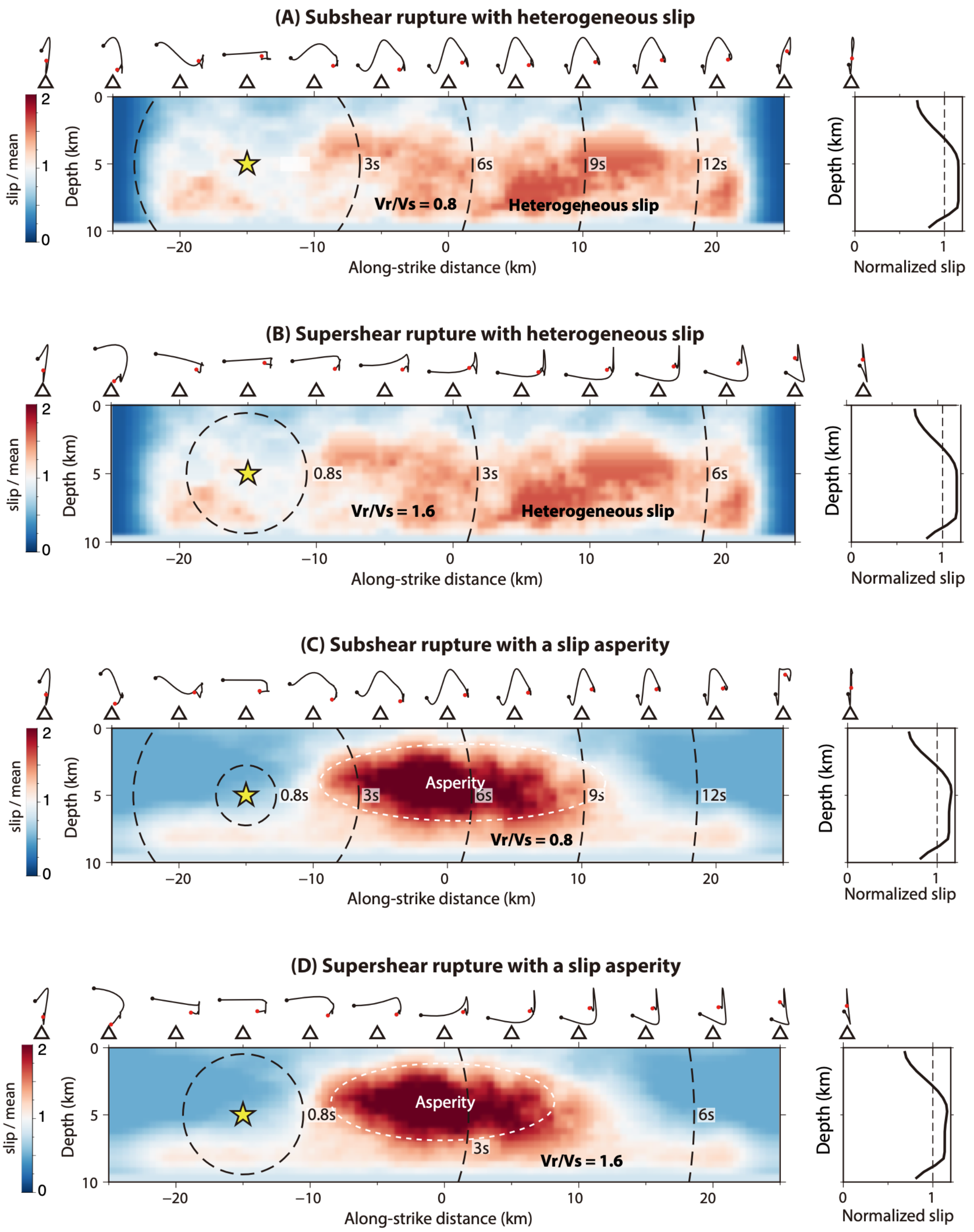

**Figure S6. Kinematic rupture models with heterogeneous slip and the resulting near-fault particle motions. (A and B)** Subshear and supershear rupture models with small-scale heterogeneities. **(C and D)** Subshear and supershear rupture models with a large slip asperity. The small panels on the right are the depth distributions of normalized cumulative slip in these models.

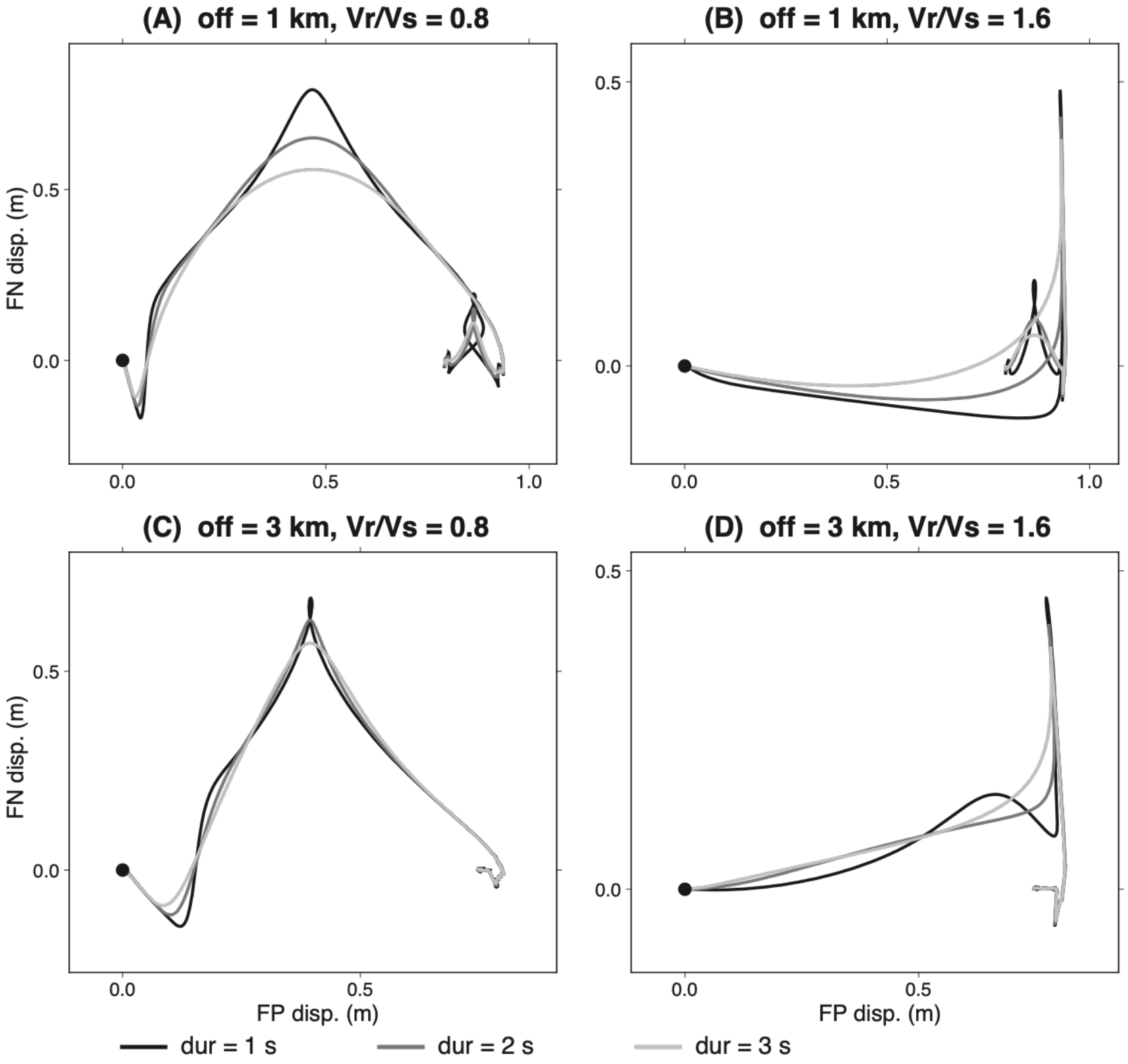

**Figure S7. Particle motions with different slip pulse durations at different off-fault distances and rupture speeds.** The synthetics with durations of 1 s, 2 s, and 3 s are plotted in black, dark gray, and light gray, respectively.

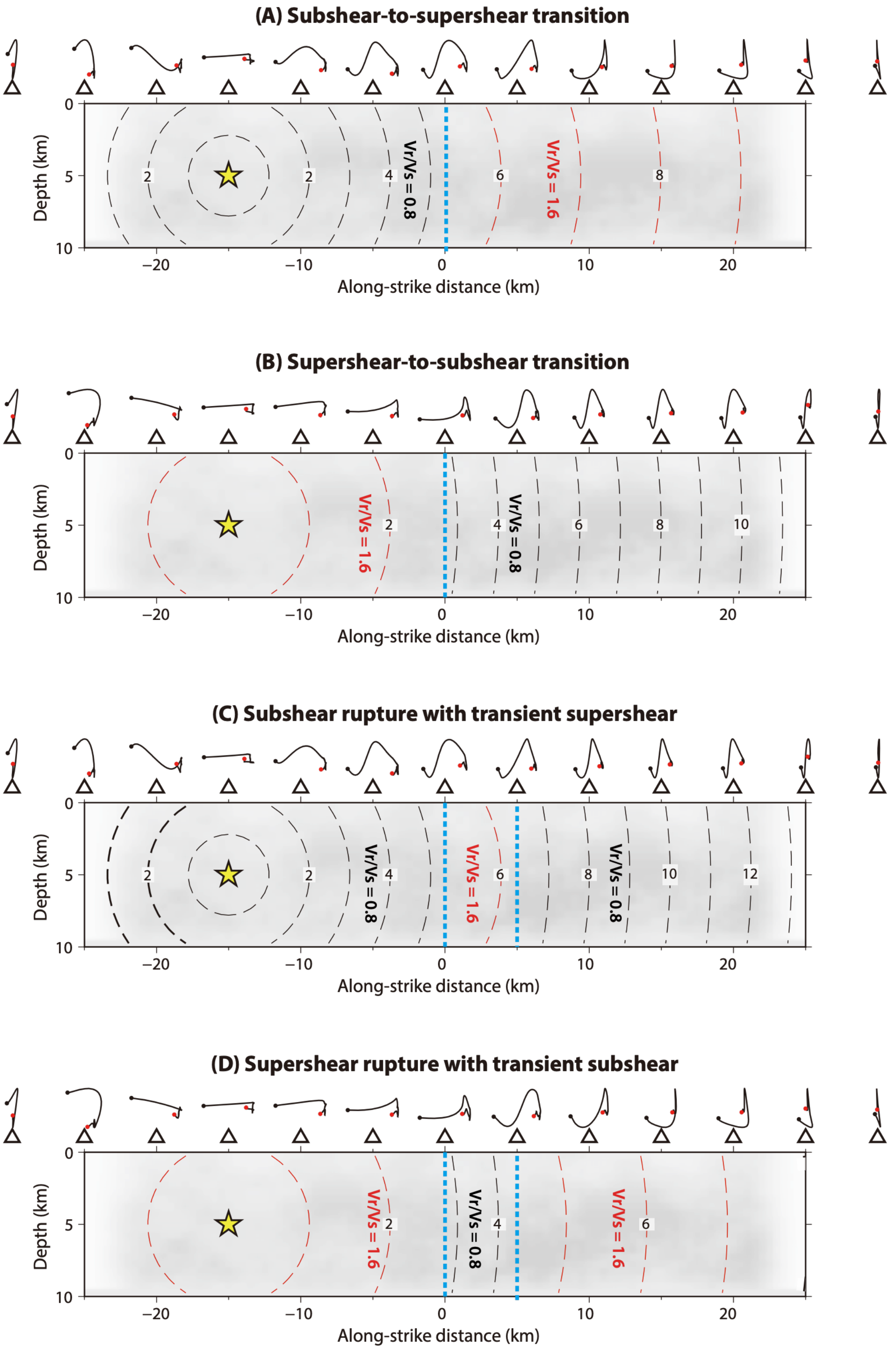

**Figure S8. Particle motions with rupture speed transitions. (A and B)** Rupture models with a subshear-to-supershear and a supershear-to-subshear transition, respectively. The blue dashed lines mark the location where the transitions occur. **(C and D)** Rupture models with transient supershear and subshear propagation, respectively. The blue dashed lines indicate the location of the transient zones. Particle motions along the fault at an off-fault distance of 1 km are plotted at the top of each panel.

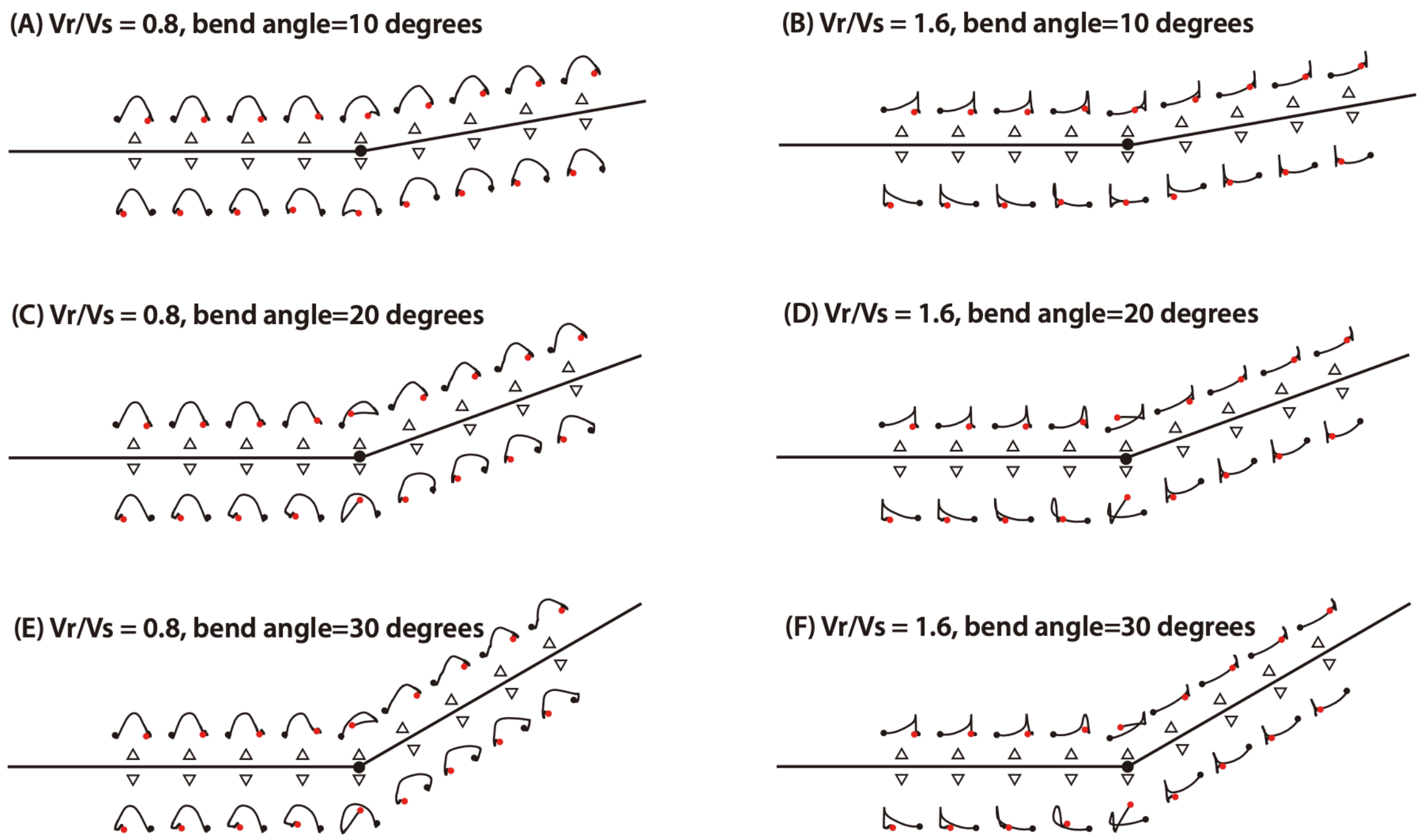

**Figure S9. Particle motions near fault bends. (A and B)** Particle motions at an off-fault distance of 1 km along the fault near a bend of 10° at subshear and supershear rupture speeds, respectively. **(C and D)** Near a bend of 20°. **(E and F)** Near a bend of 30°. The ruptures propagate from left to right. The length of each segment is 40 km.

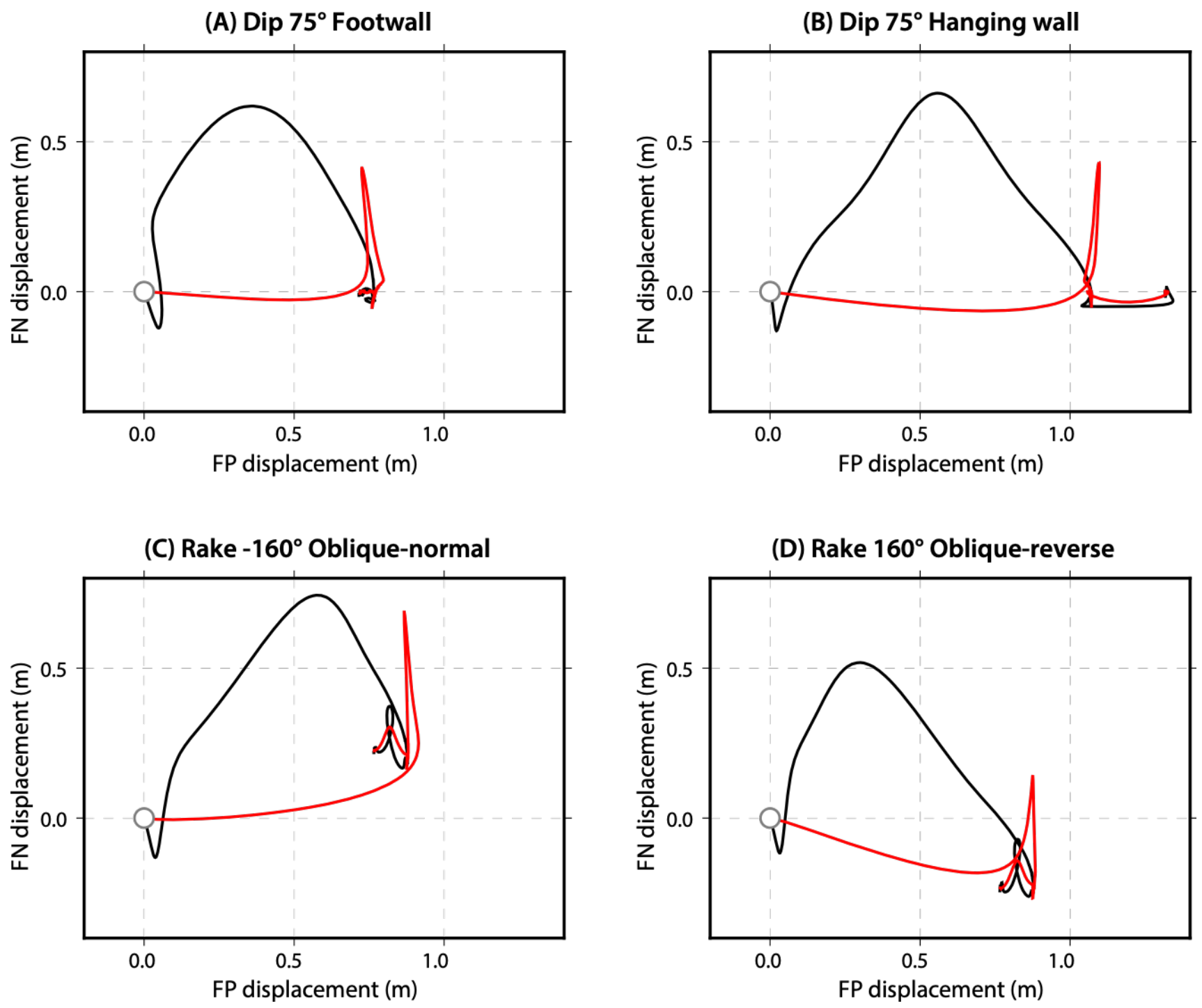

**Figure S10. Influence of dip and rake angles on particle motion.** Particle motions under subshear (0.8*Vs) and supershear (1.6*Vs) are plotted in black and red respectively.

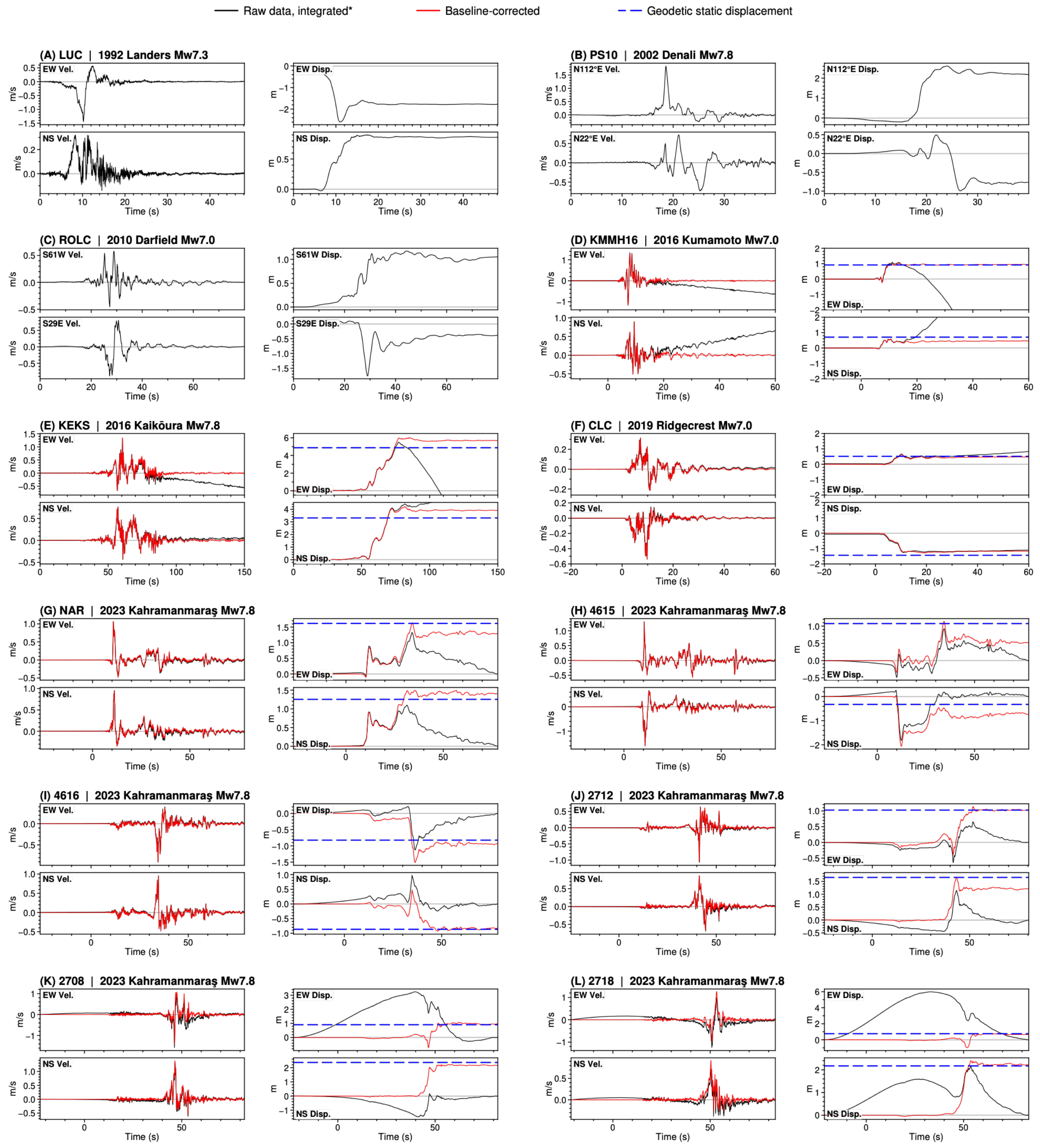

**Figure S11. Baseline correction results.** For each station, the two left panels show the raw (black) and baseline-corrected (red) velocity waveforms in the two horizontal components, and the two right panels show the corresponding displacement waveforms. The blue dashed lines indicate the static displacements from geodetic observations (*9–13*). The records at LUC and PS10 had already been baseline-corrected in previous studies, and the records at ROLC show no obvious baseline shift. The records at these three stations are therefore not corrected.

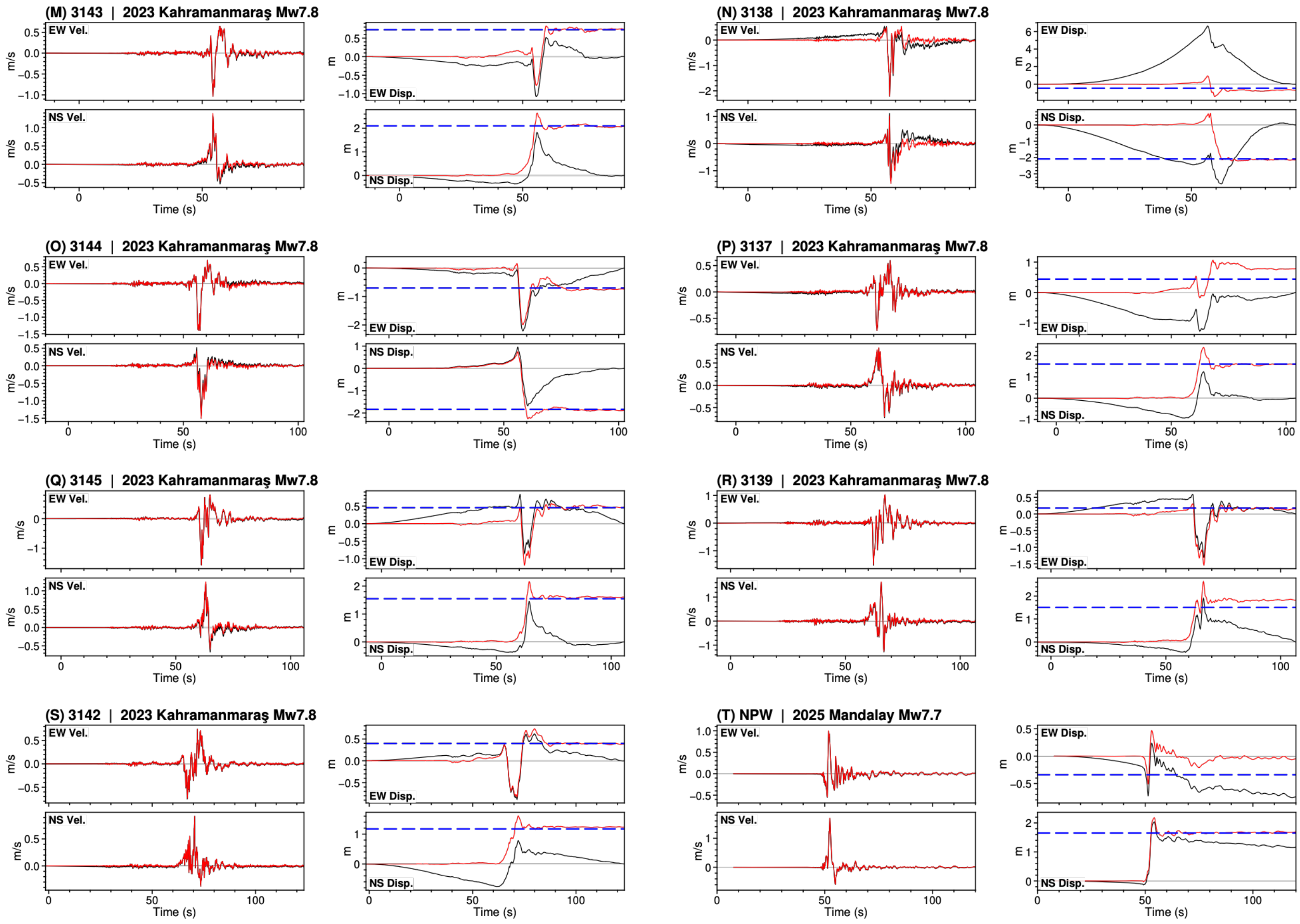

**Figure S11 (continued)**

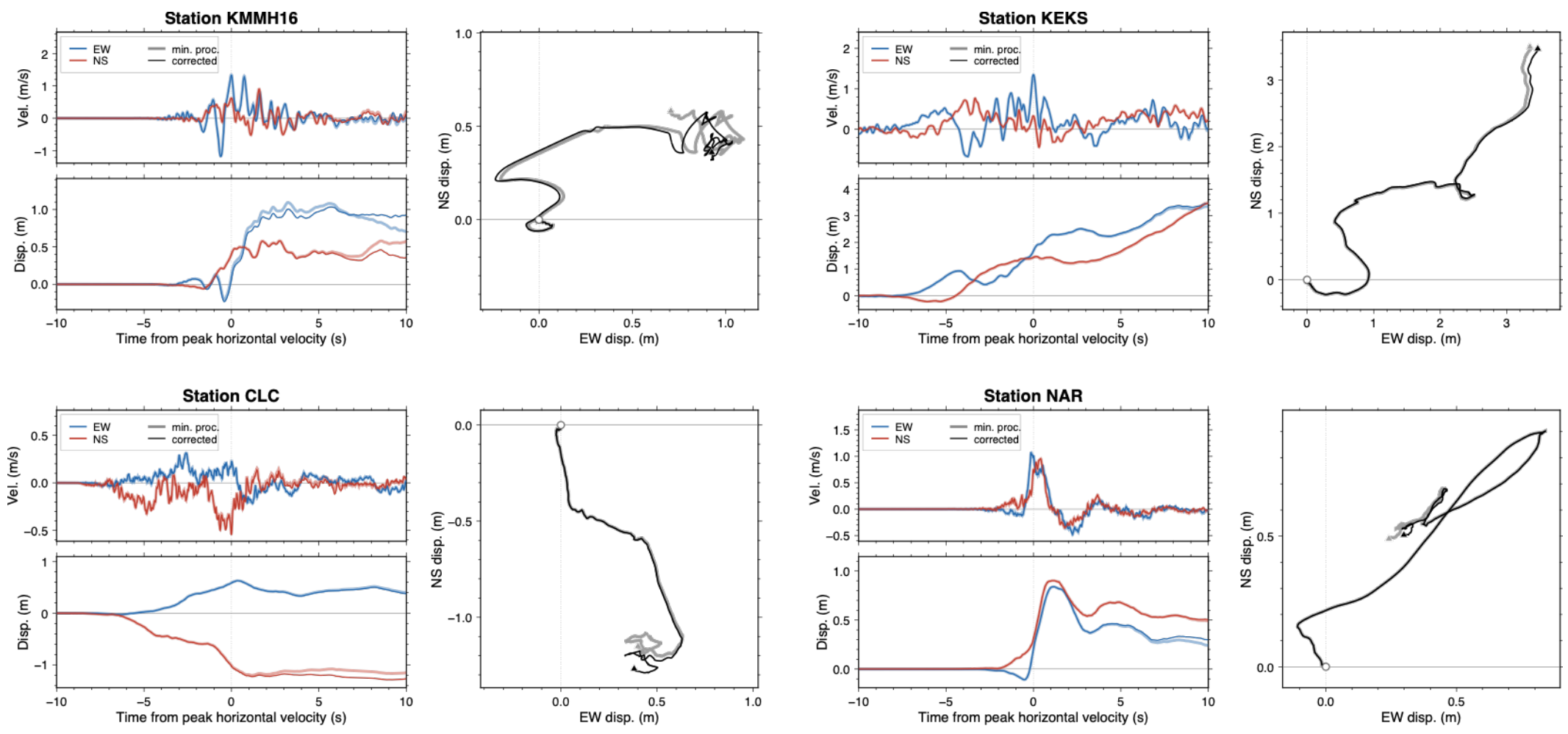

**Figure S12. Comparison of minimally processed data and fully corrected data during the large velocity pulses.** For each station, the left two panels show the horizontal velocity and displacement waveforms in a 20s-long window centered at the peak horizontal velocity. The displacements are calculated only based on the velocity waveforms in this time window. The right panel show the comparison of particle motions in this time window.

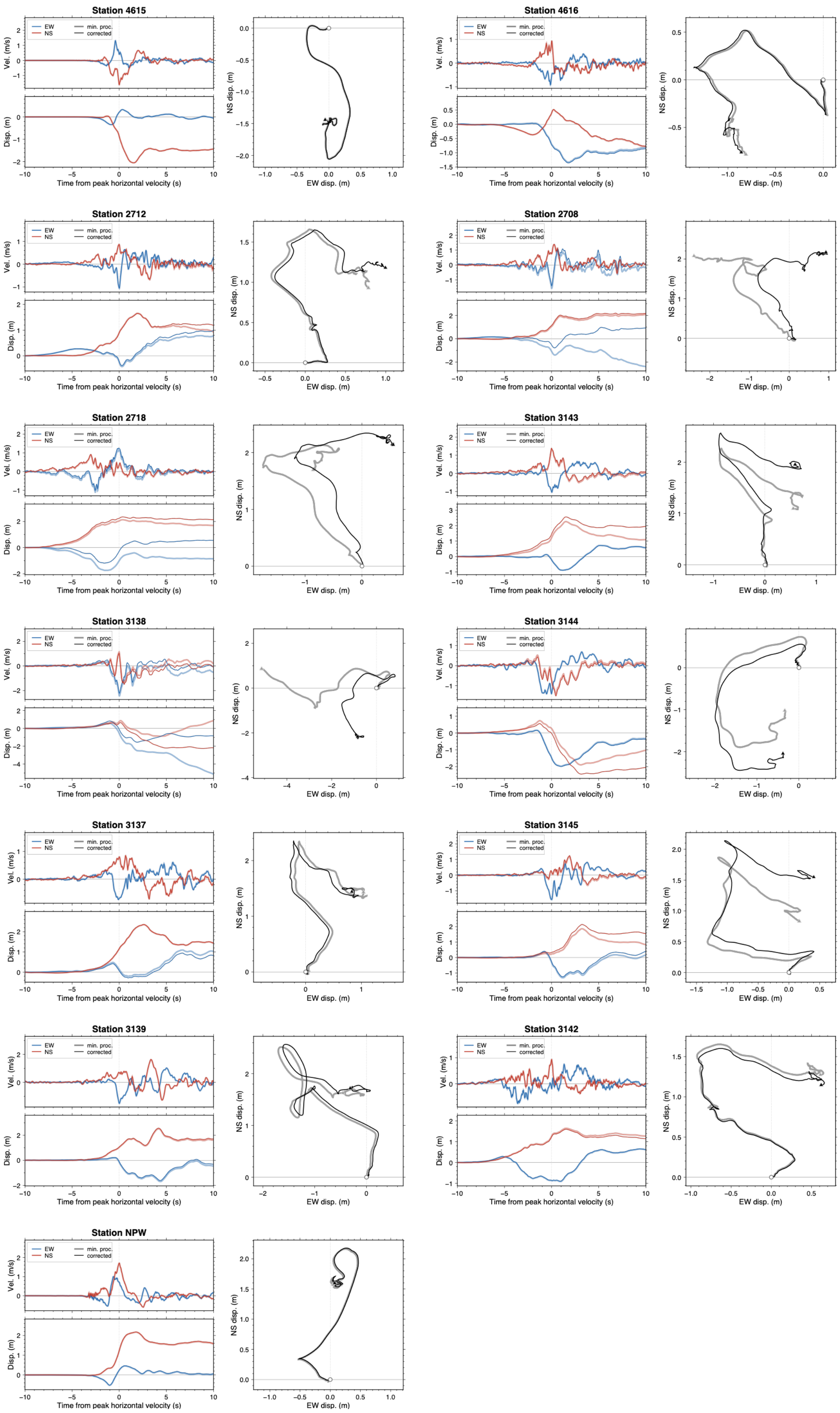

Figure S12 (continued)

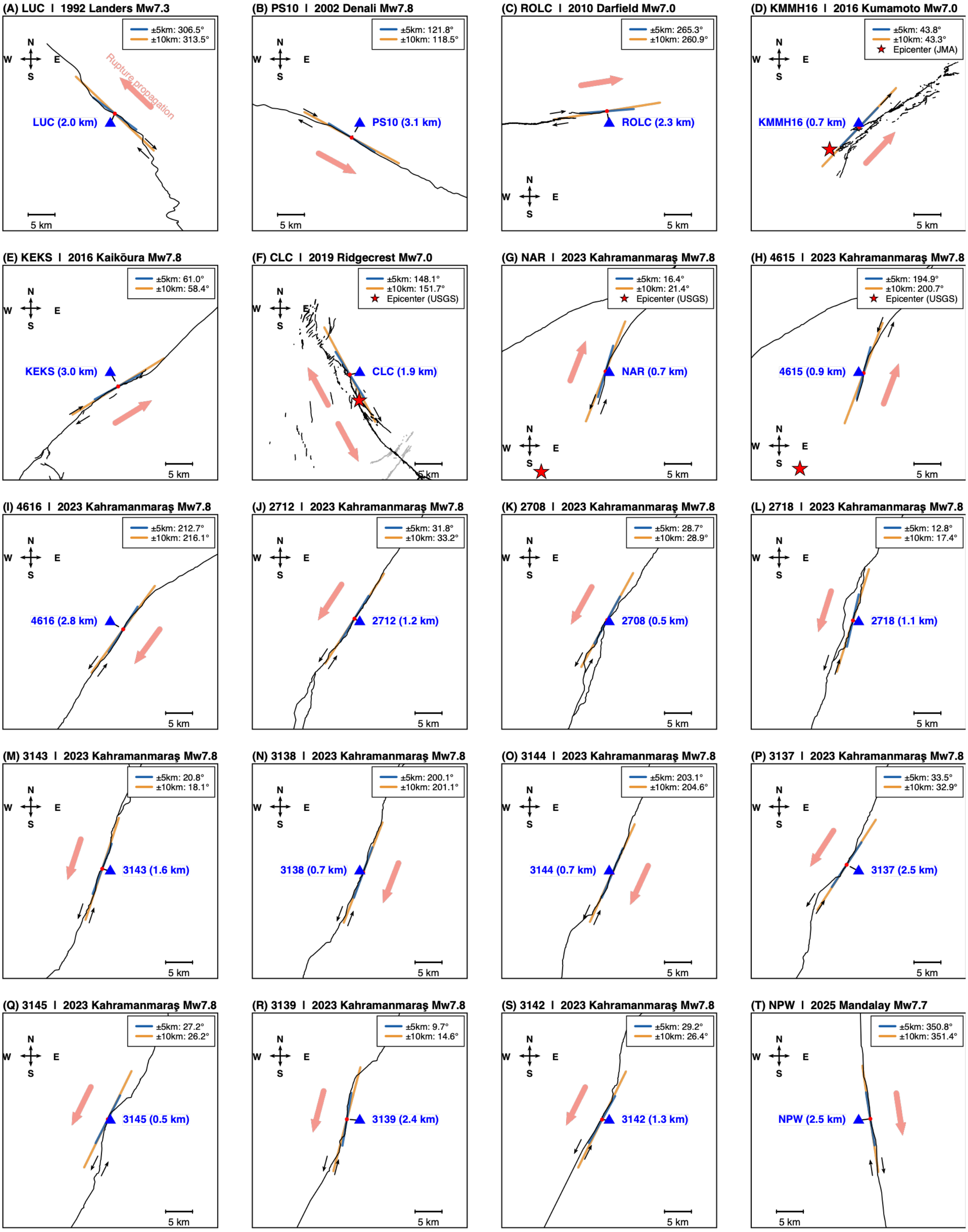

**Figure S13. Near-fault stations (blue triangles) and local surface rupture traces (black lines).** The distances to the fault traces are written in blue. The blue and orange lines represent the estimated strikes using the surface traces within 5 km and 10 km to the stations, respectively. The strike angles are written in the upper right corners. The thick light red arrows indicate the rupture propagation direction. The small black arrows mark the slip directions.

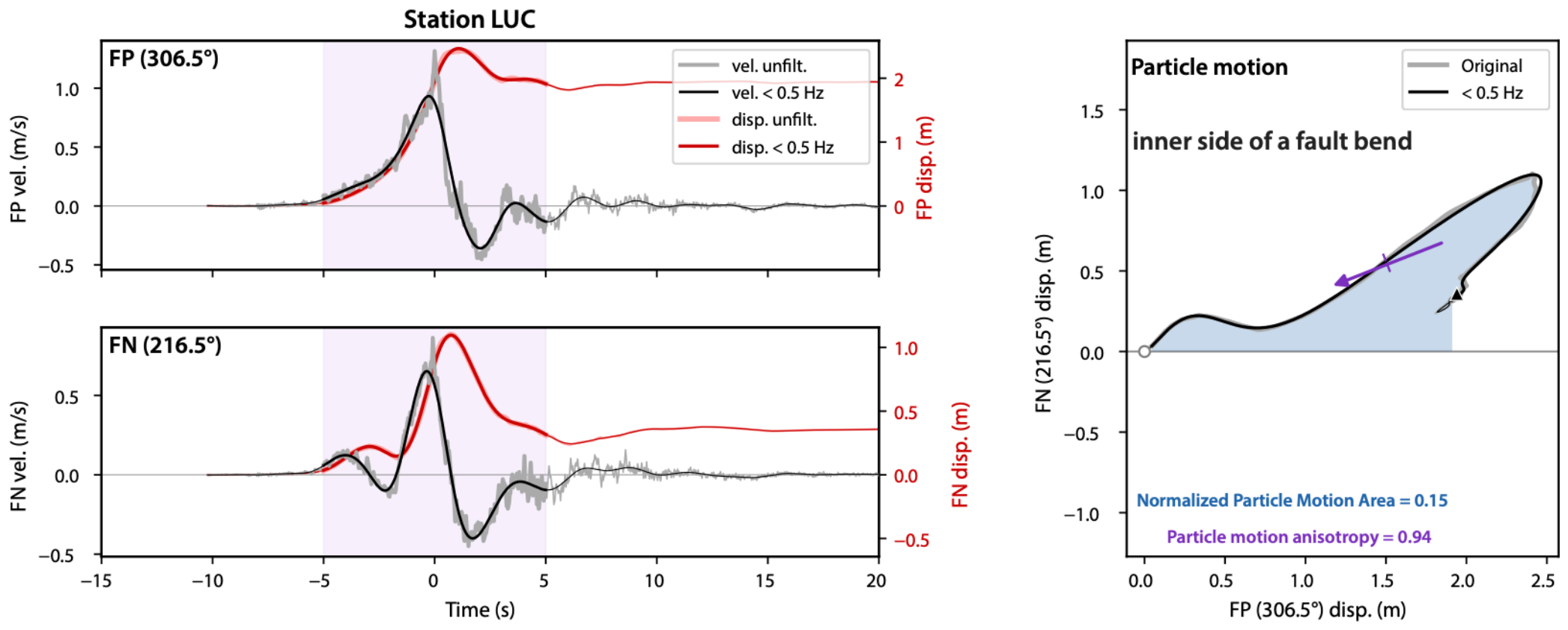

**Figure S14. Velocity and displacement waveforms (left) and particle motion (right) at station LUC in the 1992 Mw 7.3 Landers earthquake.** The raw velocity and displacement waveforms are plotted in gray and light red, respectively, and the waveforms low-pass filtered at 0.5 Hz are plotted in black and red, respectively. The purple shading marks the window used for the anisotropy analysis. In the particle-motion panel, the motions before and after filtering are plotted in gray and black, respectively. The particle motion area (FN > 0) is shaded in blue. The purple arrow and the dashed line indicate the major and minor axes in the anisotropy analysis. Waveforms and particle motions inside and beyond the analysis window are plotted with thick and thin lines, respectively. LUC is in the inner side of a bend. Therefore, the rupture speed is inconclusive based on the particle motion trend.

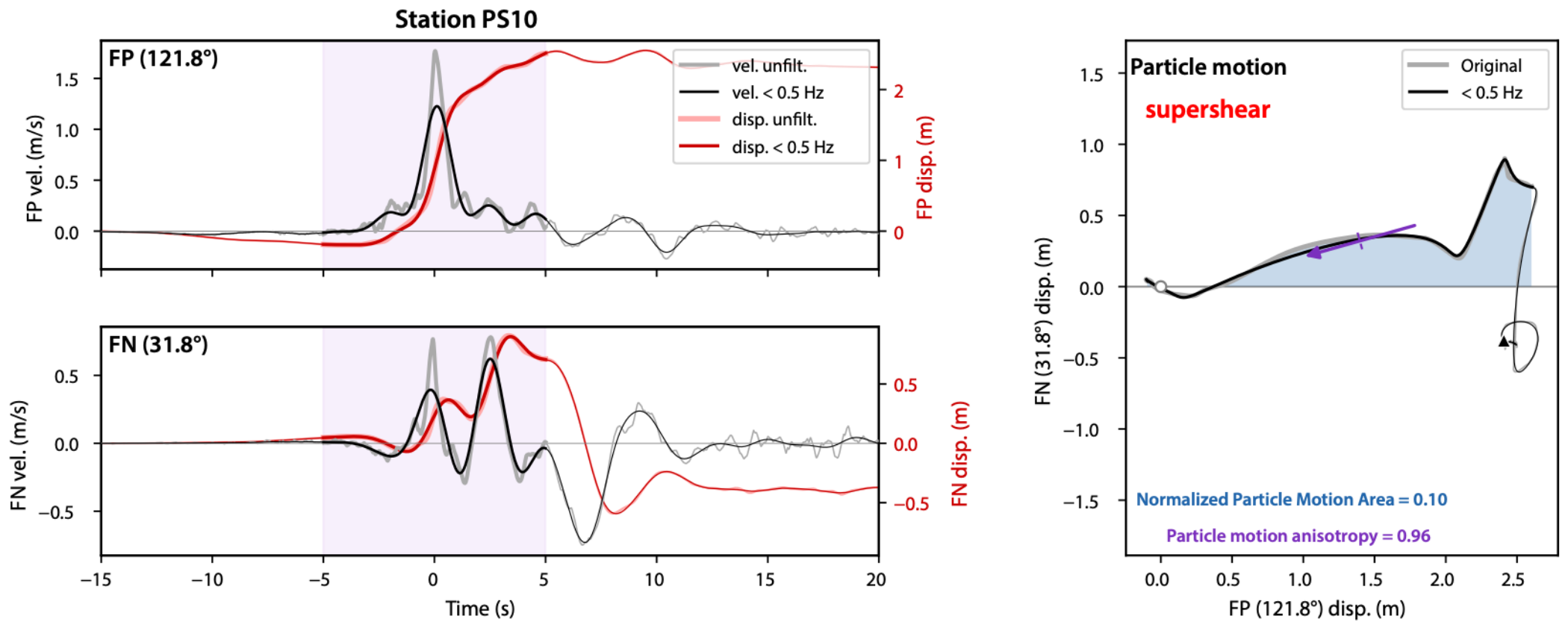

**Figure S15. Same as Figure S14 but for station PS10 in the 2002 Mw 7.8 Denali earthquake.** The particle motion at PS10 indicates supershear rupture.

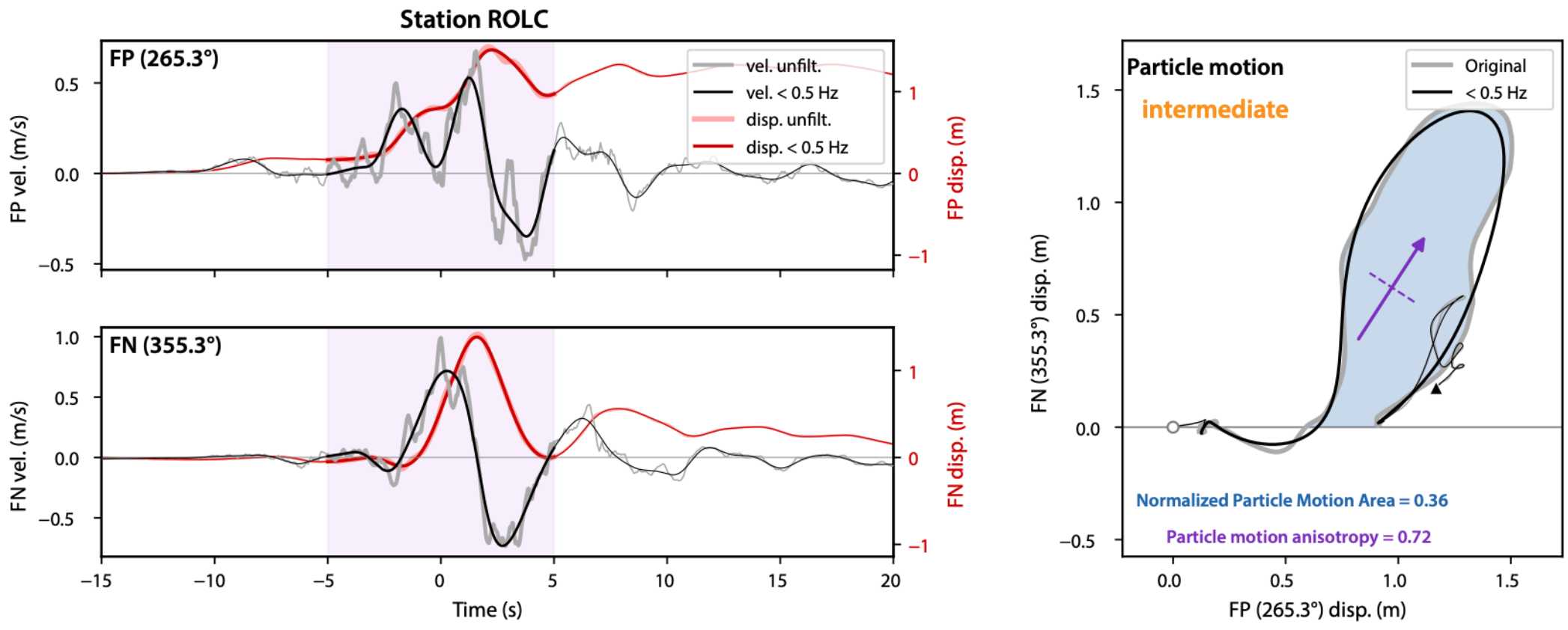

**Figure S16. Same as Figure S14 but for station ROLC in the 2010 Mw 7.0 Darfield earthquake.** The particle motion at ROLC indicates an intermediate rupture speed.

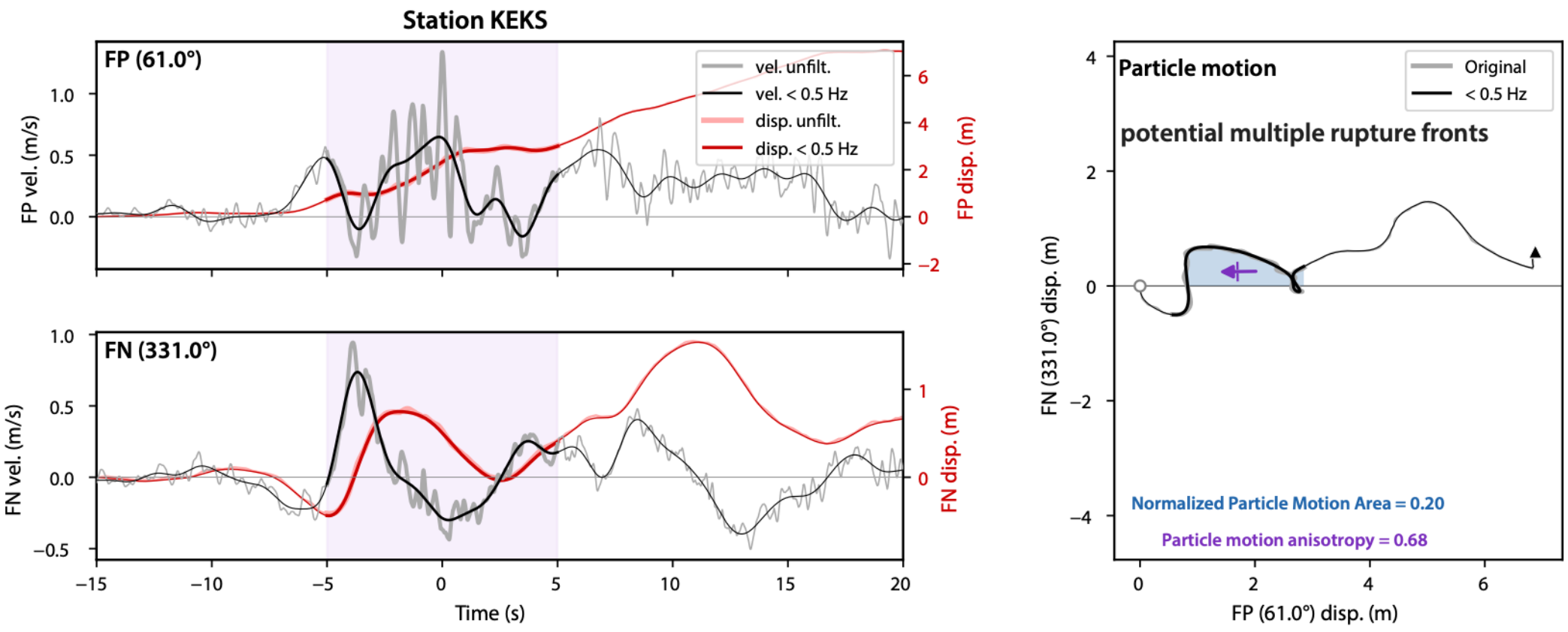

**Figure S17. Same as Figure S14 but for station KEKS in the 2016 Mw 7.8 Kaikōura earthquake.** The particle motion at KEKS indicates a complex local rupture with potential multiple rupture fronts.

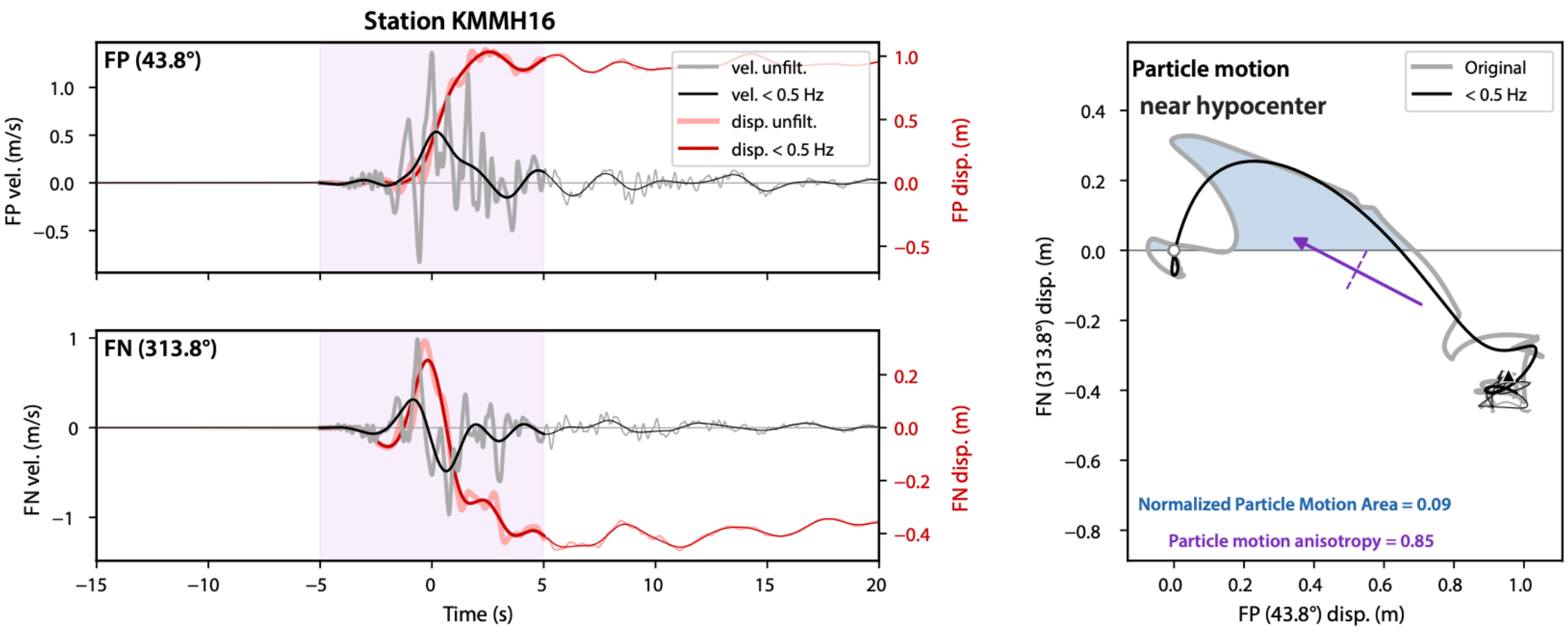

**Figure S18. Same as Figure S14 but for station KMMH16 in the 2016 Mw 7.0 Kumamoto earthquake.** Because KMMH16 is located close to the hypocenter, its particle motion cannot constrain the rupture speed.

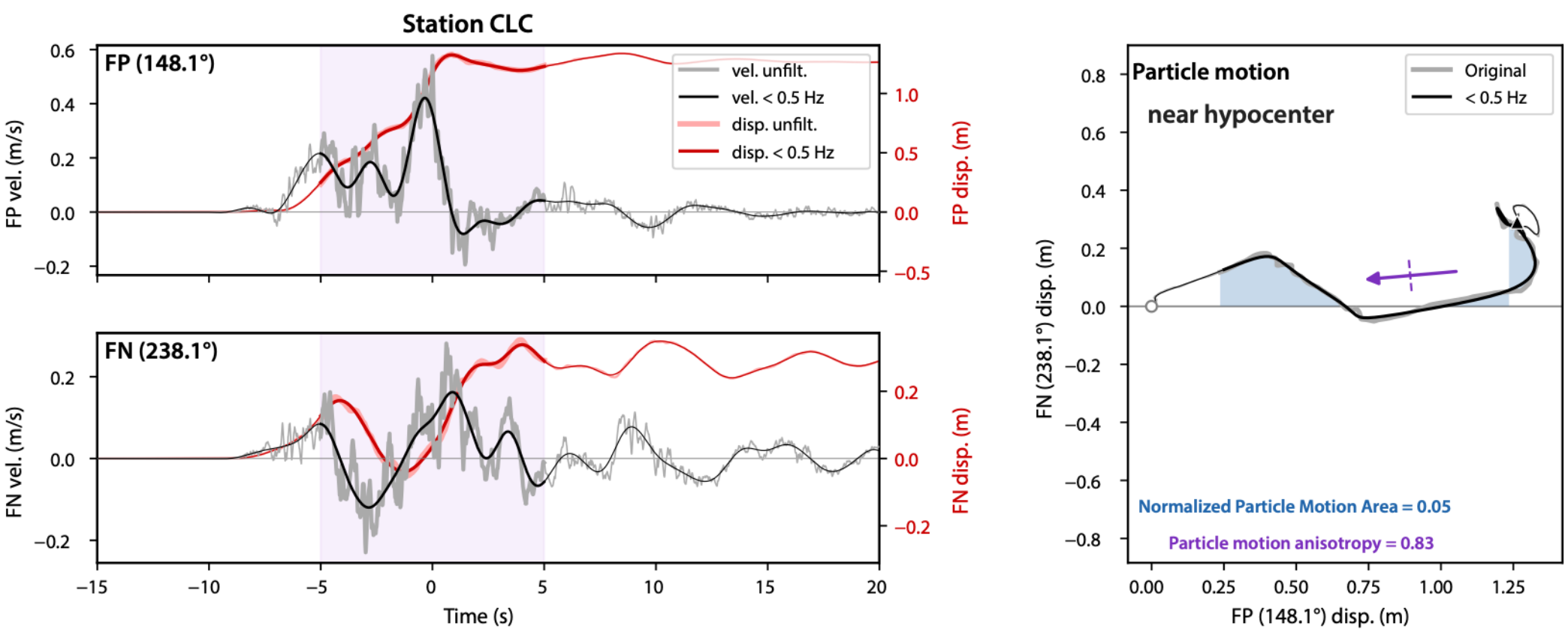

**Figure S19. Same as Figure S14 but for station CLC in the 2019 Mw 7.0 Ridgecrest earthquake.** Because CLC is located close to the hypocenter, its particle motion cannot constrain the rupture speed.

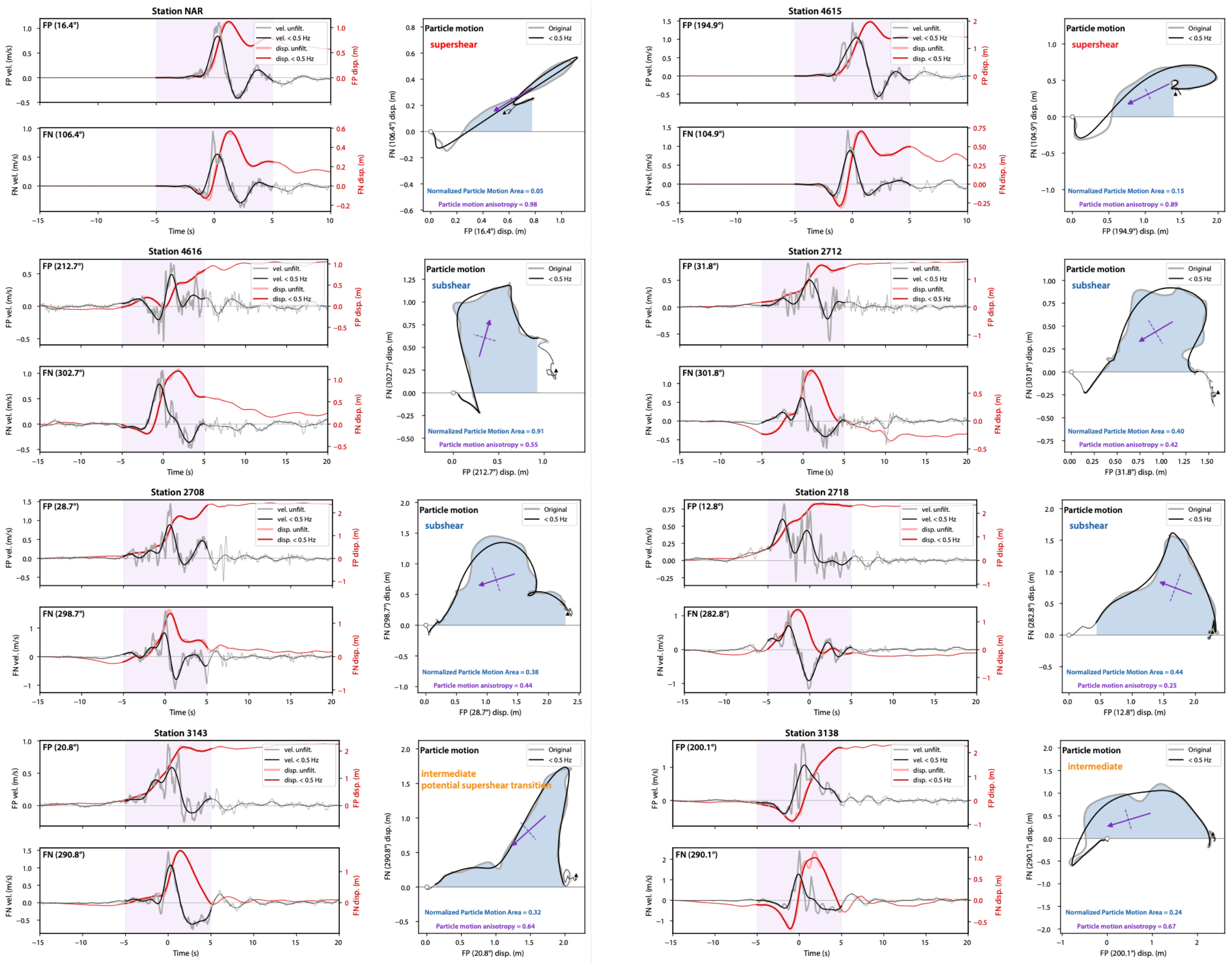

**Figure S20. Same as Figure S14 but for near-fault stations in the 2023 Mw 7.8** Kahramanmaraş **earthquake.**

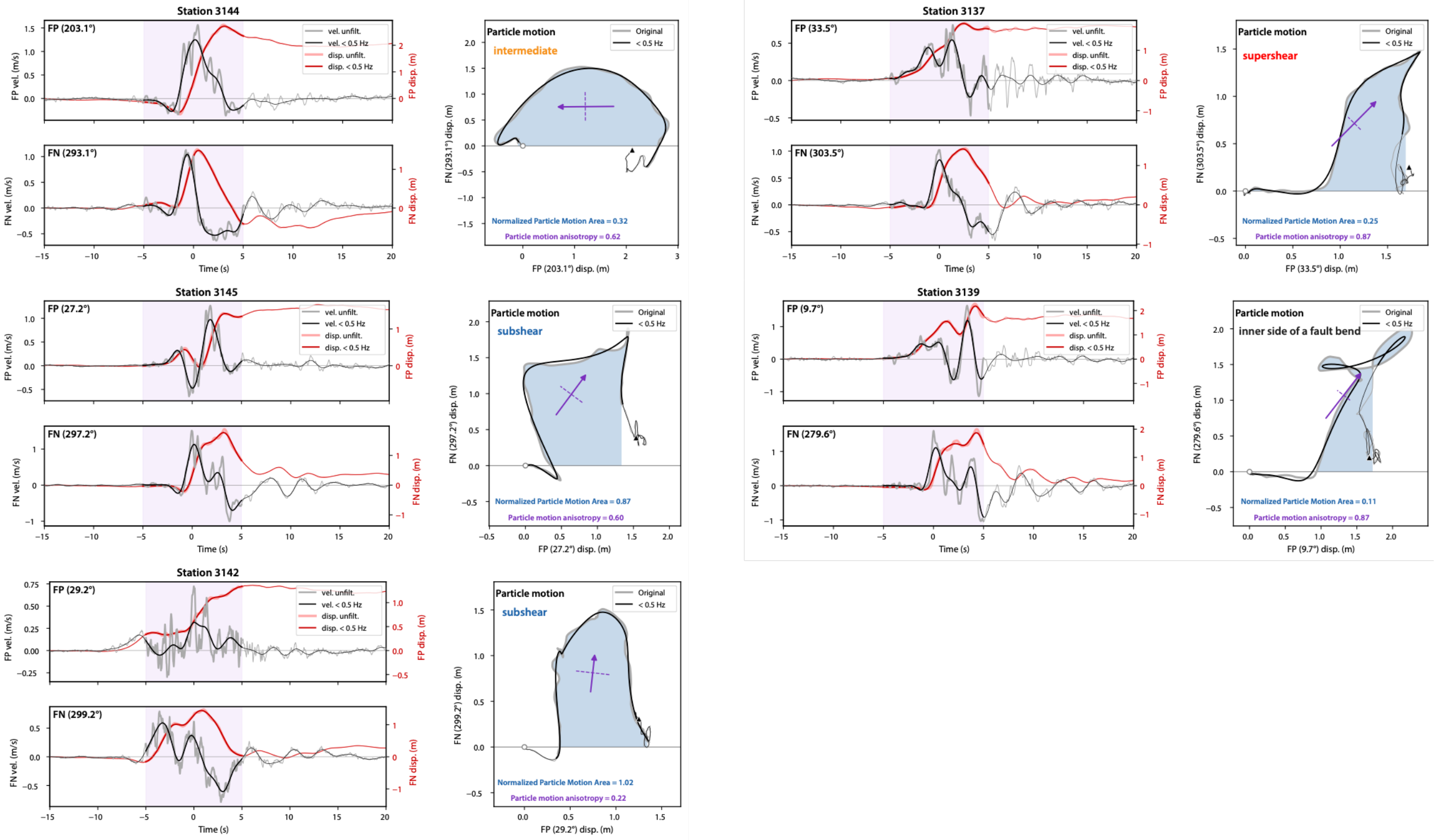

# Figure S20 (continued)

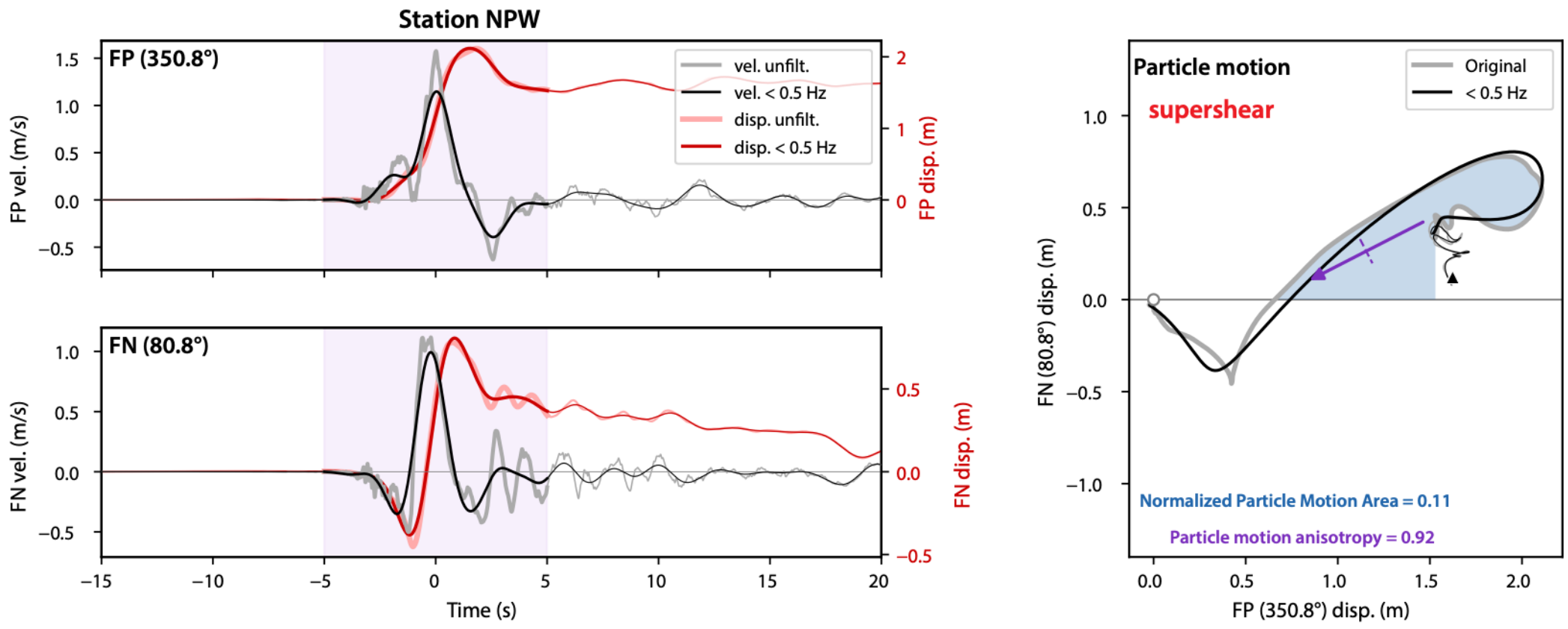

**Figure S21. Same as Figure S14 but for station NPW in the 2025 Mw 7.7 Mandalay earthquake.** The particle motion at NPW indicates supershear rupture.

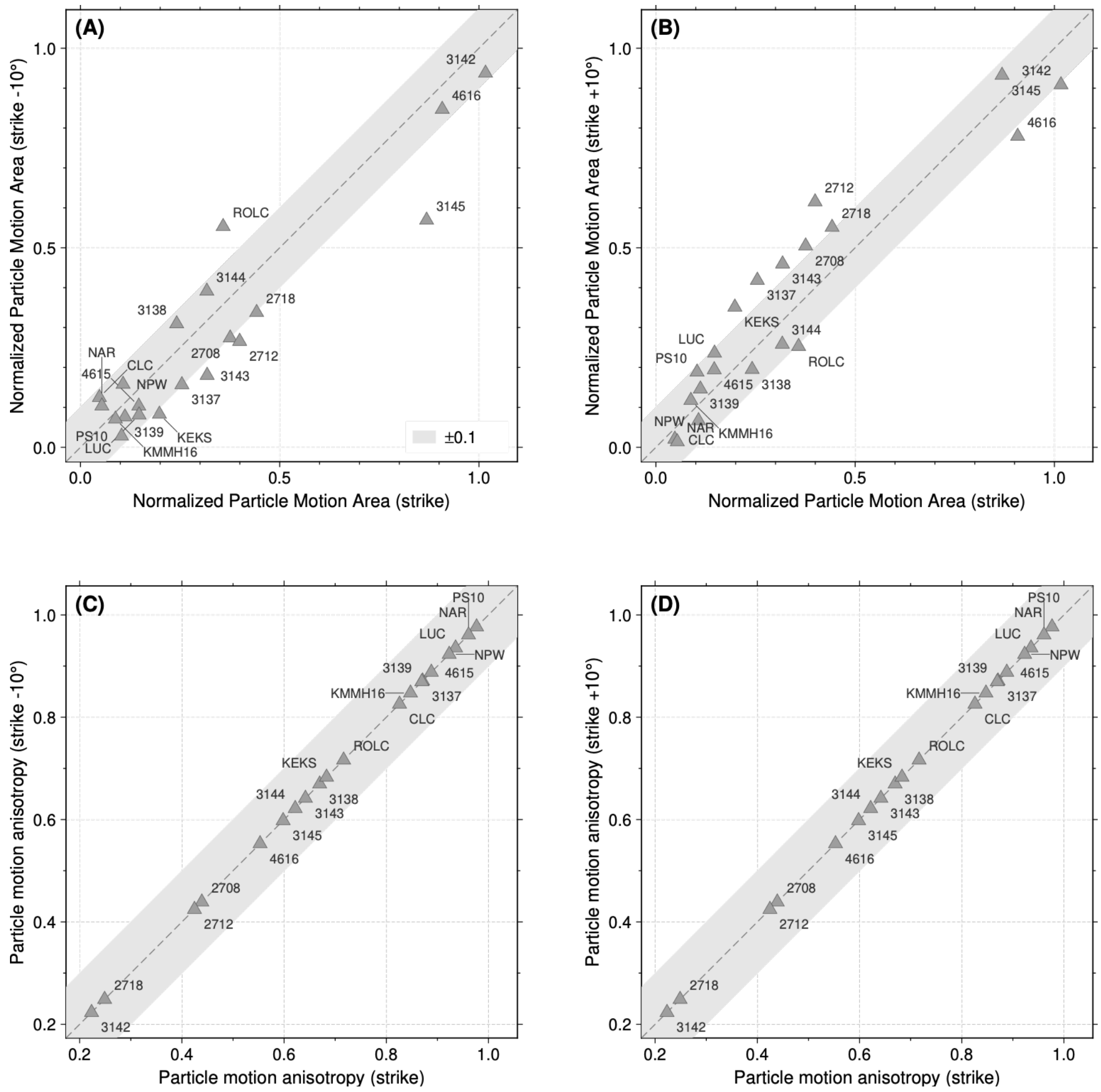

**Figure S22. Influence of the local strike on the shape metrics. (A and B)** Normalized particle motion areas computed with the local strike rotated by $-10°$ and $+10°$ from the estimated local strike, respectively. **(C and D)** Same as (A) and (B) but for the particle motion anisotropy. The gray bands indicate deviations of $\pm 0.1$.

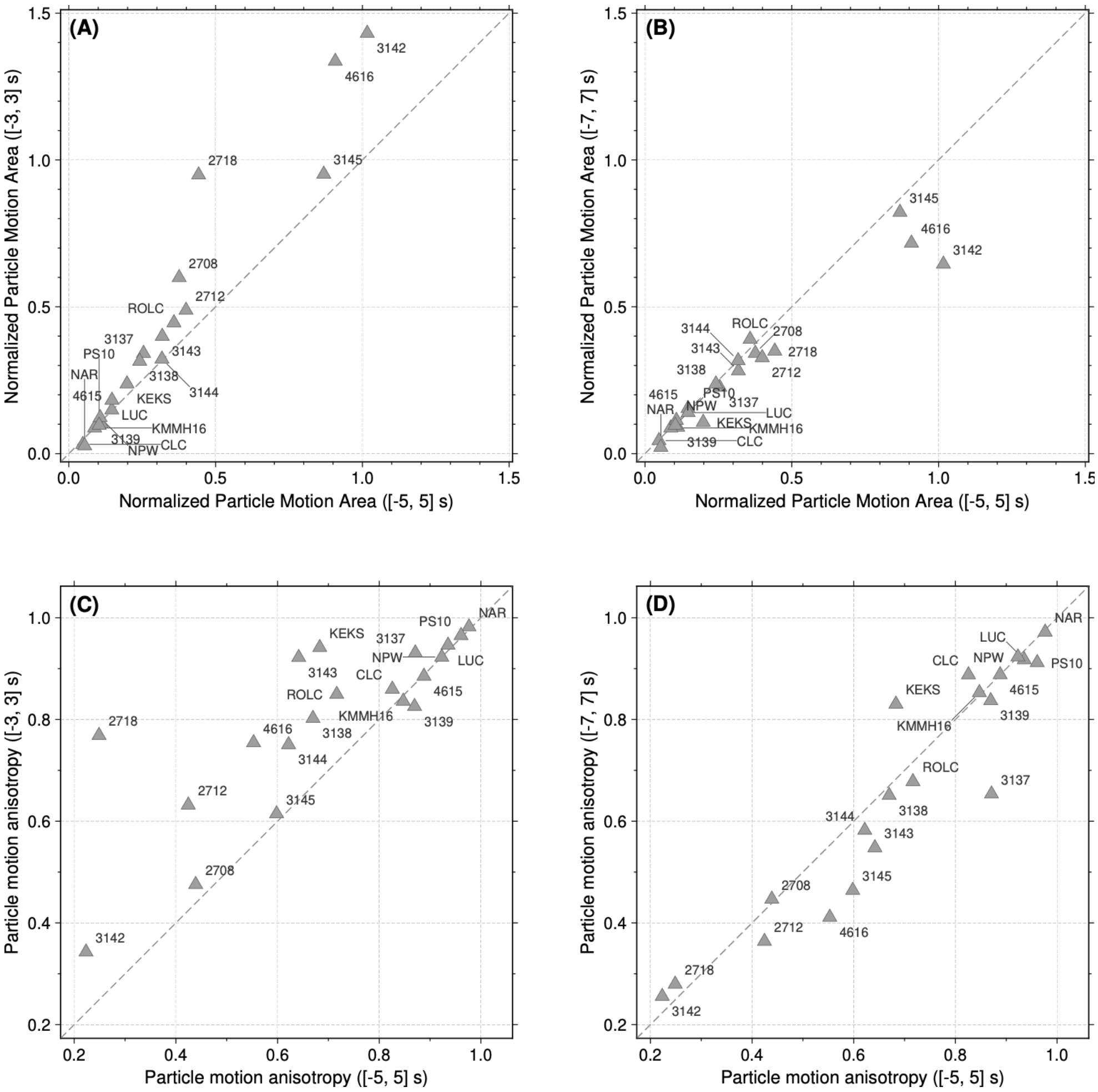

**Figure S23. Influence of the time window on the shape metrics. (A and B)** Normalized particle motion areas computed with time windows of ±3 s and ±7 s relative to the peak horizontal velocity, respectively, compared with the results using a window of ±5 s. **(C and D)** Same as (A) and (B) but for the particle motion anisotropy.